\documentclass[aps,prd,superscriptaddress,twocolumn,nofootinbib]{revtex4-2}
\usepackage{graphicx}
\usepackage[caption=false]{subfig}
\usepackage{epstopdf}
\usepackage{amsmath}
\usepackage{amsfonts}
\usepackage{amssymb}
\usepackage{latexsym}
\usepackage{hyperref}
\usepackage[english]{babel}
\usepackage[utf8]{inputenc}
\usepackage[colorinlistoftodos]{todonotes}
\usepackage{color}
\usepackage{slashed}
\usepackage{feynmp}
\usepackage{bm}
\usepackage{bbold}
\usepackage{eufrak}
\usepackage{slashed}
\usepackage{bm}  
\usepackage{xcolor}
\usepackage{footnote}
\usepackage{multirow}

\usepackage{ulem}

\begin{document}
	
\title{Self-dual Maxwell-Chern-Simons solitons in a parity-invariant scenario}

\author{W. B. De Lima}
\email{wellissonblima@cbpf.br}
\affiliation{Centro Brasileiro de Pesquisas F\'{i}sicas (CBPF), Rua Dr Xavier Sigaud 150, Urca, Rio de Janeiro, Brazil, CEP 22290-180}
	
\author{P. De Fabritiis}
\email{pdf321@cbpf.br}
\affiliation{Centro Brasileiro de Pesquisas F\'{i}sicas (CBPF), Rua Dr Xavier Sigaud 150, Urca, Rio de Janeiro, Brazil, CEP 22290-180}

		

\begin{abstract}
We present a self-dual parity-invariant  $U(1) \times U(1)$ Maxwell-Chern-Simons scalar $\text{QED}_3$. We show that the energy functional admits a Bogomol'nyi-type lower bound, whose saturation gives rise to first order self-duality equations. We perform a detailed analysis of this system, discussing its main features and exhibiting explicit numerical solutions corresponding to finite-energy topological vortices and non-topological solitons. The mixed Chern-Simons term plays an important role here, ensuring the main properties of the model and suggesting possible applications in condensed matter.
\end{abstract}


\maketitle
\pagestyle{myheadings}



\section{Introduction} \label{sec_intro}

It is well-known that the Abelian-Higgs (AH) model in 2+1 dimensions admits finite-energy, electrically neutral, topological vortices with a quantized magnetic flux~\cite{NielsenOlesen}. These can be seen as the relativistic generalization of the Abrikosov vortices~\cite{Abrikosov} in the Ginzburg-Landau model~\cite{GinzburgLandau}.
Remarkably, there is a special coupling regime in which these vortices satisfy first order equations~\cite{Bogomol'nyi}, allowing one to find exact vortex solutions in the AH model~\cite{VegaSchaposnik}.

In the presence of a Chern-Simons (CS) term~\cite{DeserJackiwTempleton1, DeserJackiwTempleton2,Schonfeld, ChernSimons,Dunne}, the property of flux attachment allows the existence of electrically charged vortices with finite energy~\cite{PaulKhare}, and in the pure CS limit, peculiar charged vortices with finite energy were also shown to exist~\cite{JatkarKhare}. 
Self-dual vortices, satisfying first order equations coming from the saturation of a Bogomol'nyi-type lower bound for the energy, were found both in the pure CS limit~\cite{HongKimPac, JackiwWeinberg, JackiwLeeWeinberg} and in the Maxwell-CS model~\cite{LeeLeeMin, Kim2, Rim}.
It is well-known that self-duality is closely related with $\mathcal{N}=2$ supersymmetry~\cite{Susy1, Susy2}, and vortices were also studied in this scenario~\cite{Susy3,Susy4,Susy5, Susy6}.
For interesting reviews, see Refs.~\cite{Dunne2, Horvathy1}.
 

It is commonly accepted that a CS term implies parity  and time-reversal violation, but this is not always true, as was already pointed out in the pioneering works~\cite{DeserJackiwTempleton1,DeserJackiwTempleton2}, and explicitly shown a decade later~\cite{Hagen}(see also~\cite{Wilczek}). This discussion had as a background experiments suggesting parity-invariance in high-$T_c$ superconductors\cite{Keifl, Spielman, Lyons}, motivating the development of theoretical models for superconductivity agreeing with these results~\cite{Semenoff, Mavromatos,  Kovner, Mavromatos2}.  

Vortices in similar contexts were already discussed in the literature before~\cite{Kim, Diziarmaga1, Diziarmaga2, Shin1, Shin2}, but without Maxwell terms.
In a previous work~\cite{PippoWell}, we studied vortices in a parity-invariant Maxwell-CS model, filling a gap in the vortex literature.
Interestingly, similar models with a mixed CS term find many applications in condensed matter~\cite{Kou1, Kou2, Kou3, Kou4, Qi, Ye, Diamantini1, Diamantini2, Diamantini3, Sakhi, Oswaldo, Wellisson}.



In this work, we present a self-dual version of the parity-invariant Maxwell-CS model~\cite{PippoWell}. We show that by choosing a suitable potential, the model admits a Bogomol'nyi-type bound for the energy, whose saturation leads to first order self-duality equations. We perform a detailed analysis of this system, examining its main properties. We present explicit numerical solutions corresponding to finite-energy topological vortices and non-topological solitons, discussing their main features.


This work is organized as follows: in Sec.~\ref{theoretical_setup} we build the main theoretical setup; in Sec.~\ref{self_duality}, we present the self-duality equations and discuss the boundary conditions; we exhibit and discuss explicit numerical solutions in Sec.~\ref{explicit_solutions}; finally, in Sec.~\ref{conclusion} we state our conclusions.
We adopt $\epsilon^{0 1 2} = -1$ and $\epsilon^{i j} \equiv \epsilon^{0 i j}$, for the Levi-Civita tensor, with Latin indices referring to spatial components.


\section{Theoretical setup}\label{theoretical_setup}

Let us consider the parity-invariant $U(1)_A \times U(1)_a$ Maxwell-CS  model coupled with scalar matter:
\begin{align}\label{MainLagrangian}
	\mathcal{L} = &-\frac{1}{4} F_{\mu \nu} F^{\mu \nu} -\frac{1}{4} f_{\mu \nu} f^{\mu \nu}  + \mu \epsilon^{\mu \nu \rho} A_\mu \partial_\nu a_\rho  \nonumber \\
	&+ \vert D_\mu \phi_+ \vert^2 + \vert D_\mu \phi_- \vert^2   +  \frac{1}{2} \left(\partial_\mu N\right)^2 + \frac{1}{2} \left(\partial_\mu M\right)^2 \nonumber \\
	&- V\left(\vert \phi_+ \vert, \vert \phi_- \vert, N, M\right),
\end{align}
where we define $D_\mu \phi_\pm = \partial_\mu \phi_\pm +i e A_\mu \phi_\pm \pm i g a_\mu \phi_\pm$.
Here we have $F_{\mu \nu} = \partial_\mu A_\nu - \partial_\nu A_\mu$ and $f_{\mu \nu} = \partial_\mu a_\nu - \partial_\nu a_\mu $, the gauge couplings associated with $U(1)_A$ and $U(1)_a$ are $e$ and $g$, respectively, and $\mu >0$ is the CS parameter.
	
This model has $U(1)_A \times U(1)_a$ gauge symmetry by construction, with the following transformations:
 \begin{align}  	
  	\begin{array}{l l}		
  \left\{    \begin{array}{l l} 
  	\phi_{\pm}'     &  =e^{i\rho{(x)} }\phi_\pm ,\\     \noalign{\bigskip}
  	A'_\mu & =A_\mu-\frac{1}{e}\partial_\mu\rho{(x)}~, \\ \noalign{\bigskip}
  	a'_\mu & =a_\mu ~;  
  \end{array} \right.   	& \left\{    \begin{array}{l l} 
  \phi_{\pm}'     &  =e^{\pm i\xi{(x)} }\phi_\pm ,\\     \noalign{\bigskip}
  A'_\mu & =A_\mu~, \\ \noalign{\bigskip}
  a'_\mu & =a_\mu -\frac{1}{g}\partial_\mu\xi{(x)} ~.  
\end{array} \right.    
  	\end{array}	
\end{align}
   
Parity invariance is achieved here by extending the usual parity concept to include a transformation that swaps the role of $\phi_+$ and $\phi_-$ in the space of fields:
\begin{align}\label{paritytransf}
	A_\mu^P &= \mathcal{P}_\mu^{\; \nu } \, A_\nu, \quad a_\mu^P = - \mathcal{P}_\mu^{\; \nu} \, a_\nu, \nonumber \\
	\phi_+^P &= \eta \, \phi_-, \quad \,\,\,\,\, \phi_-^P = \eta \, \phi_+, \nonumber \\
	N^P &= N, \quad \quad \,\,\, M^P = -M,
\end{align}
where $\mathcal{P}_\mu^{\; \nu} = diag(+-+)$, and $\eta$ is a complex phase. 
This model is also time-reversal invariant, but we will not discuss it here, since we will be mainly interested in static configurations. 
Here one could also define~\cite{PippoWell} $A_\mu^{\pm} = \left(A_\mu \pm a_\mu\right)/\sqrt{2}$, with parity acting as $A_\mu^\pm \rightarrow \mathcal{P}_\mu^{\nu} A_\nu^\mp$.


To the purpose of investigating the existence of self-dual solitons, let us propose the following potential:
\begin{align}\label{SelfDualPotential}
	V &=  (eN+gM)^2 \vert \phi_+ \vert^2 +(eN-gM)^2 \vert \phi_- \vert^2 \nonumber\\
	& +\frac{1}{2}  \left[e\left(\vert \phi_+ \vert ^2 -\vert \phi_- \vert ^2\right) -\mu M\right]^2\nonumber\\
	& +\frac{1}{2}  \left[g\left(\vert \phi_+ \vert ^2 +\vert \phi_- \vert ^2  -2v^2\right)  -\mu N\right]^2.
\end{align}
This potential is consistent with all the symmetries of the model, and despite not being the most general possibility, it arises naturally from the requirement of a Bogomol'nyi bound for the energy.
Setting $\vert \phi_+ \vert = \vert \phi_-\vert$ , $M=0$, $e=g$, and appropriately rescaling the remaining parameters, it exactly reproduces the potential proposed in \cite{LeeLeeMin}, which is known to contain both pure Maxwell and pure CS self-dual vortices as limiting cases. Instead, if $N = M = 0$ and $e = g$, we can recover the potential used in Ref.~\cite{PippoWell}.

The equations of motion for this model are given by
\begin{align}\label{EOM}
	\partial_\mu F^{\mu \nu} + \mu \epsilon^{\nu \alpha \beta} \partial_\alpha a_\beta &= e \left(J_+^\nu + J_-^\nu\right), \nonumber \\
	\partial_\mu f^{\mu \nu} + \mu \epsilon^{\nu \alpha \beta} \partial_\alpha A_\beta &= g \left(J_+^\nu - J_-^\nu\right), \nonumber \\		
	D_\mu D^\mu \phi_\pm &= -\frac{dV}{d\phi_\pm^*},
\end{align}
supplemented by the equations for $N$ and $M$
\begin{align}\label{eomN}
	(\Box + \mu^2)N &= \! -2e \left[(eN + gM)\vert \phi_+ \vert^2 +(eN - gM)\vert \phi_- \vert^2\right]    \nonumber \\
	&+\mu \left[g (\vert \phi_+ \vert^2 + \vert \phi_- \vert^2 - 2v^2)\right] \nonumber \\
	(\Box + \mu^2)M &= \! -2g \left[(eN + gM)\vert \phi_+ \vert^2 -(eN - gM)\vert \phi_- \vert^2\right]  \nonumber \\
	&+\mu \left[ e (\vert \phi_+ \vert^2 - \vert \phi_- \vert^2)  \right].
\end{align}
The currents above are $J_\pm^\nu = i \left[ \phi_\pm^* D^\nu \phi_\pm - \phi_\pm D^\nu \phi_\pm^*  \right]$. 
We can define the electric and g-electric fields as $E^i = F^{i 0}$, $e^i = f^{i 0}$ as well as the magnetic and g-magnetic fields as $B = \epsilon^{i j} \partial_i A_j$, $b = \epsilon^{i j} \partial_i a_j$. From the zeroth component of the gauge fields equations, we can obtain
\begin{align}\label{GaussLaw}
	\vec{\nabla} \cdot \vec{E} + \mu b = e \left(\rho_+ + \rho_- \right), \nonumber \\
	\vec{\nabla} \cdot \vec{e} + \mu B = g \left(\rho_+ - \rho_- \right),
\end{align}
where $\rho_\pm = J^0_\pm$. The electric and g-electric charges are given by $Q = e \int \! d^2x \, \left(\rho_+ + \rho_-\right)$, $G = g \int \! d^2x \, \left(\rho_+ - \rho_-\right)$, and the magnetic and g-magnetic fluxes by $\Phi \equiv  \int \! d^2x \,  B $, $\chi \equiv  \int \! d^2x \,  b$, respectively. Upon integration, we obtain:  
\begin{align}\label{key}
	Q = \mu \chi, \quad G = \mu \Phi,
\end{align}
relating the charge associated with one gauge field with the magnetic flux associated with the other. This mutual statistics behavior~\cite{Wilczek} is a characteristic feature of models with a mixed CS term~\cite{Kim},  here implementing the flux attachment in a parity-invariant way.

The energy functional for this model is of the form:
\begin{align}\label{EnergyFunctional}
	E \!=\!\! \int \! &d^2x \Bigg[ \frac{1}{2}\left(\vec{E}^2 + B^2\right) + \frac{1}{2}\left(\vec{e}^2 + b^2\right) +  V  \nonumber \\
	&+ \vert D_0\phi_+ \vert^2  + \vert D_0\phi_- \vert^2 +  \vert D_i\phi_+ \vert^2 + \vert D_i\phi_- \vert^2         \nonumber \\
	&+ \frac{1}{2}\Big((\partial_0 M)^2 \!+\! (\partial_0 N)^2 \!+\! (\partial_i M)^2  \!+\! (\partial_i N)^2\Big)\Bigg].
\end{align}
The minimum energy configuration can be achieved, for instance, with $A_\mu = a_\mu = 0$ and constant scalar fields, provided that they minimize the potential. Inspection of Eq.~\eqref{SelfDualPotential} indicates that $V=0$ if, and only if:
\begin{align}
	&(eN+gM)^2 \vert \phi_+ \vert^2  = 0, \nonumber\\
	&(eN-gM)^2 \vert \phi_- \vert^2  = 0, \nonumber \\
	&e\left(\vert \phi_+ \vert ^2 -\vert \phi_- \vert ^2\right) -\mu M = 0, \nonumber\\
	&g\left(\vert \phi_+ \vert ^2 +\vert \phi_- \vert ^2  -2v^2\right)  -\mu N = 0.
\end{align}
Out of which only four possibilities arise:
\begin{align}
	&\bullet (0,0):  \vert \phi_+ \vert^2  \!=\! \vert \phi_- \vert^2 \!=\! 0; M \!=\! 0; N \! = \! -\frac{2gv^2}{\mu}, \nonumber \\
	&\bullet (1,1): \vert \phi_+ \vert^2  \!=\! \vert \phi_- \vert^2 \!=\! v^2 ; M \!=\! N \!=\! 0, \nonumber \\
	&\bullet (0,1): \vert \phi_+ \vert^2  \!=\! 0; \vert \phi_- \vert^2  \!=\! v^2; M \!=\! -\frac{ev^2}{\mu}; N \!=\! -\frac{gv^2}{\mu},\nonumber \\
	&\bullet (1,0): \vert \phi_+ \vert^2  \!=\! v^2; \vert \phi_- \vert^2 \!=\! 0;  M \!=\! \frac{ev^2}{\mu}; N \!=\! -\frac{gv^2}{\mu}. 
\end{align}

The first vacuum corresponds to the symmetric, the second, to the asymmetric, and the last two, to the partially symmetric phases of the gauge symmetry, respectively. 
Notice that the first two vacua are invariant under parity, while the last two are not, transforming into each other. 
All these vacua are degenerate, therefore, there can also be domain walls solutions connecting them, but this case will not be considered here.
We will focus on the parity-invariant vacua, investigating the existence of topological vortices and non-topological solitons.


Let us briefly discuss the perturbative spectrum by considering the quadratic part of the fluctuations around each vacuum. 
In the $(0,0)$-vacuum, one can find two massive complex scalar fields with masses $m_{\phi_+} = m_{\phi_-} = 2 e g v^2/\mu$, a real scalar and a real pseudoscalar with masses $m_N = m_M = \mu$, and two massive gauge bosons with masses equal to $\mu$. 
In the $(1,1)$-vacuum, one can find the same dispersion relation for the scalar and gauge sectors, of the form $p^2 = m^2_\pm$, with $m^2_\pm \! \!= \! \frac{1}{2} \! \left(\! \mu^2 \!+\! M_e^2 \!+\! M_g^2 \! \pm \! \sqrt{\!(\mu^2 \!+\! M_e^2 \!+\! M_g^2)^2 \!-\! 4M_e^2M_g^2}\!\right)$,
where each one has multiplicity 2 and the mass parameters are defined as $M_e^2 = 4 e^2 v^2$ and $M_g^2 = 4 g^2 v^2$. Therefore, we have 4 massive gauge and 4 massive scalar degrees of freedom with mass squared $m^2_\pm$ distributed equally.  
Here we observe pairs of propagating modes for a given mass, which is a common characteristic of self-dual models~\cite{LeeLeeMin}. The mass degeneracy for different spins even suggests a supersymmetric nature for the model.



\section{Self-duality Equations}\label{self_duality}
	
The self-dual potential~\eqref{SelfDualPotential}  makes it possible to rewrite the energy functional~\eqref{EnergyFunctional} in a very suggestive form. Upon using the equations of motion and some integrations by parts, we can write (here we are using $D_\pm \equiv D_1  \pm i D_2 $):
 \begin{align}\label{EnergyFunctional2}
	E \!=\!\! &\int \! d^2x \left[ \frac{1}{2}\left(\vec{E} \pm \vec{\nabla}{N}\right)^2  + \frac{1}{2}\left(\vec{e} \pm \vec{\nabla}{M}\right)^2 \right. \nonumber \\
	&+ \left.   \vert D_\pm\phi_+  \vert^2+  \vert D_\mp\phi_-  \vert^2  + \frac{1}{2}(\partial_0 M)^2 + \frac{1}{2}(\partial_0 N)^2\right. \nonumber \\ 
	&+\frac{1}{2}\left\lbrace B \pm \left[e\left(\vert \phi_+ \vert ^2 -\vert \phi_- \vert ^2\right) -\mu M\right] \right\rbrace^2\nonumber \\
	&+\frac{1}{2}\left\lbrace b \pm \left[g\left(\vert \phi_+ \vert ^2 +\vert \phi_- \vert ^2  -2v^2\right)  -\mu N\right] \right\rbrace^2\nonumber\\
	&   + \vert D_0 \phi_+ \mp i(eN+gM)\phi_+\vert^2 \nonumber \\
	&+  \vert D_0 \phi_- \mp i(eN-gM)\phi_-\vert^2\nonumber \\ 
	&  \left. \pm 2 g v^2 b \right].
\end{align}  
Therefore, since we have a sum of non-negative terms, we are naturally led to a  Bogomol'nyi-type bound~\cite{Bogomol'nyi} to the energy functional, given by $	E \ge 2 g v^2 \vert \chi \vert$ (for $\chi >0$ we choose the upper sign, and for $\chi <0$  we choose the lower sign).
Our main interest here is to investigate the finite-energy static field configurations that saturate these bounds, satisfying first order self-duality equations determined by Eq.~\eqref{EnergyFunctional2}.
Since we consider only the static regime, in particular, $\partial_0M = \partial_0 N = 0$.
The field configurations saturating the Bogomol'nyi bound have minimum energy given by $E = 2 g v^2 \vert \chi \vert$, and must satisfy the first order self-duality equations:
\begin{subequations}
\begin{eqnarray}
	& &D_\pm\phi_+  = 0, \quad D_\mp\phi_- = 0  \label{sd1} \\
	& &D_0 \phi_+ \mp i(eN+gM)\phi_+ = 0, \label{sd4}  \\
	& &D_0 \phi_- \mp i(eN-gM)\phi_-= 0,  \label{sd5}  \\
	& &\vec{E} \pm \vec{\nabla}{N}= 0, \quad \vec{e} \pm \vec{\nabla}{M}= 0  \label{sd6} \\
	& &B \pm \left[e\left(\vert \phi_+ \vert ^2 -\vert \phi_- \vert ^2\right) -\mu M\right]= 0, \label{sd8}  \\
	& &b \pm \left[g\left(\vert \phi_+ \vert ^2 +\vert \phi_- \vert ^2  -2v^2\right)  -\mu N\right] = 0. \label{sd9} 
\end{eqnarray}
\end{subequations}
To satisfy Eqs.~\eqref{sd4}, \eqref{sd5}, \eqref{sd6}, it is sufficient to take
\begin{align}\label{sd10}
	A_0 = \pm N, \quad	a_0 = \pm M.
\end{align}
Using the parametrization $\phi_\pm = \vert \phi_\pm \vert e^{i\omega_\pm}$ for the scalar fields, we can obtain from Eq.~\eqref{sd1} the following structure for the gauge fields spatial components:
\begin{align}\label{SelfDualGaugePotentials}
	e A_i &=\pm \frac{1}{2}\epsilon_{ij}\partial_j \ln \frac{\vert \phi_+ \vert}{v} \mp \frac{1}{2}\epsilon_{ij}\partial_j \ln \frac{\vert \phi_- \vert}{v} - \frac{1}{2}\partial_i (\omega_+ + \omega_-) \nonumber\\
	g a_i &= \pm \frac{1}{2}\epsilon_{ij}\partial_j \ln \frac{\vert \phi_+ \vert}{v} \pm \frac{1}{2}\epsilon_{ij}\partial_j \ln \frac{\vert \phi_- \vert}{v} - \frac{1}{2}\partial_i (\omega_+ - \omega_-)
\end{align}
Notice that, upon determining all the scalar fields, thanks to the self-dual structure,  we are able to obtain all the information present in the gauge fields. 
Now, acting with $\epsilon^{ki} \partial_i$, we can obtain the magnetic and g-magnetic fields.
Substituting this result into the self-dual equations \eqref{sd8} and \eqref{sd9}, after some manipulations we can find
\begin{align}\label{eqdifphi+-}
\nabla^2 \ln \frac{\vert \phi_+ \vert}{v}  &= \left[ (e^2 + g^2)\vert \phi_+ \vert ^2 -(e^2 - g^2)\vert \phi_- \vert ^2 \right. \nonumber \\
& \left. -\mu \left(e M + g N + \frac{2v^2g^2}{\mu}\right)\right] \nonumber \\
\nabla^2 \ln \frac{\vert \phi_- \vert}{v} &= \left[ (e^2 + g^2)\vert \phi_- \vert ^2 -(e^2 - g^2)\vert \phi_+ \vert ^2 \right. \nonumber \\   
&\left. -\mu \left(-e M + g N + \frac{2v^2g^2}{\mu}\right)\right].
\end{align}
Finally, from the remaining  equations, we can obtain:
\begin{align}\label{eqdifNM}
	\left(\nabla^2 \!\!- \mu^2 \right)\! N  \!&=\!  2e \left[ (eN + gM)\vert \phi_+ \vert ^2 \!+\! (eN - gM)\vert \phi_- \vert ^2  \right]   \nonumber \\
	& -\mu  \left[g\left(\vert \phi_+ \vert ^2 +\vert \phi_- \vert ^2  -2v^2\right)  \right] \nonumber \\
	\left(\nabla^2 \!\!-\mu^2\right) \! M \!&=\! 2g \left[ (eN + gM)\vert \phi_+ \vert ^2 \!-\! (eN - gM)\vert \phi_- \vert ^2  \right] \nonumber \\
	&- \mu  \left[e\left(\vert \phi_+ \vert ^2 -\vert \phi_- \vert ^2\right) \right] .
\end{align}	
This is exactly the static limit of the equations of motion for $N$ and $M$, cf. Eq.~\eqref{eomN}. The equations \eqref{eqdifphi+-} and \eqref{eqdifNM}, are the ones that must be solved in order to find the static self-dual solutions.

Let us consider here the following radially symmetric {\it ansatz} for the scalar fields:
\begin{align}\label{ScalarFieldAnsatz}
	\phi_\pm \! \left(r, \theta\right) &= v \, F_\pm (r) \, e^{i \left(m \pm n \right) \theta}, \nonumber \\
	N \! \left(r, \theta\right) &= v \, \hat{N}\left(r\right), \nonumber \\
	M \! \left(r, \theta\right) &= v \, \hat{M}\left(r\right),
\end{align}
where $m \pm n \in \mathbb{Z}$, and the profiles $F_\pm$, $\hat{N}$, and  $\hat{M}$ are dimensionless. Plugging the {\it ansatz} above in Eqs.~\eqref{SelfDualGaugePotentials}, we can obtain the gauge structure (here $\hat{\theta}_i = \epsilon_{i j} x^j /r$):
\begin{align}\label{SelfDualGaugePotentialsAnsatz}
	 A_i \! \left(r, \theta\right)  &=  \frac{1}{e r} \left[A(r) - m\right] \hat{\theta_i}, \nonumber \\
	 a_i \! \left(r, \theta\right) &= \frac{1}{g r} \left[a(r) - n\right] \hat{\theta_i}, 
\end{align}
where we defined the gauge profiles as:
\begin{align}\label{GaugeProfilesAnsatz}
	A(r) &= \pm\frac{1}{2}\left(\frac{r F'_+}{F_+} - \frac{r F'_-}{F_-}\right) \nonumber \\
	a(r) &= \pm\frac{1}{2}\left(\frac{r F'_+}{F_+} + \frac{r F'_-}{F_-}\right),
\end{align}
or, equivalently:
\begin{align}\label{GaugeProfilesAnsatz2}
	F'_+ &= \pm \, \frac{F_+ \left(A + a\right)}{r}, \nonumber \\
	F_-' &= \mp \, \frac{F_- \left(A - a\right)}{r}.
\end{align}
It should be stressed that, although the gauge field structure above \eqref{SelfDualGaugePotentialsAnsatz} is the same used in Ref.~\cite{PippoWell}, here it does not appear as an independent {\it ansatz} for the gauge fields, but it has its structure totally determined by the scalar fields {\it ansatz}, and as a consequence of the self-dual equations obtained by saturating the Bogomol'nyi bound.


Let us first discuss the profiles behavior at the origin. Looking to the gauge structure~\eqref{SelfDualGaugePotentialsAnsatz}, in order to avoid a singularity  at the origin, we must have $A(0) = m$ and $a(0) = n$. Using Eq.\eqref{GaugeProfilesAnsatz2} we see that they need to satisfy
\begin{align}
	\left( n + m \right) F_+(0) = 0, \nonumber \\
	\left( n - m \right) F_-(0) = 0.
\end{align}
These considerations imply the following behavior: 
\begin{align}\label{behaviororiginscalar}
	\left\lbrace
	\begin{array}{c}
		F_+(r) \approx r^{ \pm (n+m )}\\
		F_-(r) \approx r^{\pm( n-m)}
	\end{array}\right. ~~ \text{as}~ r \rightarrow 0.
\end{align}
Therefore, to ensure that the fields have a regular behavior at the origin, we must have $\pm n > \vert m \vert$. Notice that if we take $n=0$, we cannot ensure a regular behavior at the origin for both fields simultaneously, unless we also set $m=0$, in which case $F_+(0)$ and $F_-(0)$ remain undetermined. Finally, if we consider $n = -m \neq 0$, then $F_+(0)$ is undetermined while $F_-(0)=0$; if we consider $n=m \neq 0$, then $F_-(0)$ is undetermined while $F_+(0)=0$.
It should be noted that the behavior of $\hat{N}$ and $\hat{M}$ near the origin will follow from their equations of motion, once the behavior of $F_+$ and $F_-$ for small $r$ is determined.
	
Now, we proceed to the discussion of the asymptotic conditions. The energy contribution coming from the potential implies that for any finite-energy configurations, we must have $F_+(\infty)$ and $F_-(\infty)$ equal to $0$ or $1$. Furthermore, the covariant derivatives contribution to the energy functional include the following terms:
\begin{align}\label{key}
	E \supset 2 \pi v^2 \int dr \left[\frac{F_+^2 \left(A +a\right)^2}{r} + \frac{F_-^2 \left(A -a\right)^2}{r} \right]
\end{align}
Therefore, from the finite-energy condition, we find the following asymptotic conditions:
\begin{align}\label{key}
	\left[A(\infty) + a(\infty)\right] F_+(\infty) = 0, \nonumber \\
	\left[A(\infty) - a(\infty)\right] F_-(\infty) = 0. 
\end{align}	
First of all, let us consider the case in which the scalar profiles asymptote to the $(1,1)$-vacuum, that is, when $F_+(\infty) = F_-(\infty) = 1$. In this case, we are dealing with topological vortices, and we must have  $A(\infty) = a(\infty) = 0$. These configurations have quantized magnetic fluxes ($\Phi = \frac{2 \pi}{e} \, m$ and $\chi = \frac{2 \pi}{g} \, n$), charges ($Q = \frac{2 \pi}{g} \mu n $ and $ G = \frac{2 \pi}{e} \mu m $) and energy ($ E = 2 g v^2 \vert \chi \vert = 4 \pi  v^2 \vert n \vert $).

Furthermore, we consider the case in which we asymptote to the $(0,0)$-vacuum, that is, when we have $F_+(\infty) = F_-(\infty) = 0$. In this case, we can generically assume that we have $F_+(r) \approx 1 / r^{\pm\left(\alpha + \beta\right)}$ and $F_-(r) \approx 1 / r^{\pm\left(\alpha - \beta\right)}$ in the limit $r \rightarrow \infty$, with $\pm\alpha > \beta$. We are now dealing with non-topological vortices, and we have $A(\infty) \equiv - \beta$ and $a(\infty) \equiv -\alpha$. These configurations do not have quantized magnetic fluxes, that now are given by $\Phi = \frac{2 \pi}{e} (m + \beta)$ and $\chi = \frac{2 \pi}{g} (n + \alpha)$, being $\alpha $ and $\beta $ real numbers. Nonetheless, they are also self-dual configurations that still saturate the energy bound, that can be written as $E = 4 \pi  v^2 \vert n + \alpha \vert$.
We remark that the case $\alpha =0$ only makes sense if we also take $\beta = 0$, recovering therefore the previous situation of topological vortices.

Last but not least, we comment that there is another asymptotic behavior that could be considered. One could also analyze finite-energy configurations that asymptote to a parity-breaking vacuum by considering $F_+(\infty) = 0, \, F_-(\infty) = 1$ or $F_+(\infty) = 1, \, F_-(\infty) = 0$, which amounts to choosing $\beta = \alpha$ or $\beta = -\alpha$, respectively. These will not be investigated here, because we are only concerned with the parity-preserving scenario.

The angular momentum of these configurations is given by $J = \int \! d^2x \, \epsilon^{ij} r_i T_{0j}$.
In Ref.~\cite{PippoWell}, one can find the following expression for the angular momentum of the finite-energy, static, rotationally symmetric vortices: $J = \frac{2 \pi \mu}{e g} \left[A(0) a(0) - A(\infty) a(\infty)\right]$.
It should be noted that the model considered here does not lead to any change in the above angular momentum expression.
Here we have $A(0)=m, a(0)=n$ and also $A(\infty) \equiv -\beta, a(\infty) \equiv -\alpha$. Thus, we can rewrite this expression in the following way:
\begin{align}
J = \frac{2 \pi \mu}{e g} \left(n \, m -  \alpha \, \beta \right) = \frac{Q G}{2 \pi \mu} - \frac{Q}{e} \beta - \frac{G}{g} \alpha.
\end{align} 
This is in agreement with the result found in Ref.\cite{Kim}. 



\section{Explicit solutions and discussion}\label{explicit_solutions}

\begin{table}[]
	\renewcommand{\tabcolsep}{4.5 pt}
	\renewcommand{\arraystretch}{1.4}
	\begin{tabular}[t]{|c|c|c|c|c|c|c|c|}
		\hline
		\multirow{2}{*}{$(n,m)$} &
		\multicolumn{2}{c|}{$(\frac{1}{2},\frac{1}{2})$} &
		\multicolumn{2}{c|}{$(1,0)$} & 
		\multicolumn{2}{c|}{$(\frac{3}{2},\frac{1}{2})$} &
		$(0,0)$ \\ 
		& NT & T & NT & T & NT & T & NT \\ \hline
		$E(1/4\pi v^2)$ & 5.38 & 0.50 & 5.76 & 1.00 & 8.53 & 1.50  & 3.25 \\
		\hline
		$J(eg/2\pi\mu)$ & -4.78 & 0.25 & 0.81& 0.00 & 0.75 & 0.75  & 0.75 \\
		\hline
		$\Phi (e/2\pi)$ & 1.53 & 0.50 & -0.17  &0.00 & 0.50 & 0.50 &  -0.23  \\
		\hline
		$\chi (g/2\pi)$ & 5.38 & 0.50 & 5.76 & 1.00 & 8.53 & 1.50 & 3.25 \\
		\hline
	\end{tabular}
	\caption{Physical properties of topological vortices (T) and non-topological solitons (NT) for different values of $n$ and $m$.}
	\label{tableTNT}
\end{table}

In this section, we exhibit explicit numerical solutions for the self-duality equations. First of all, we rewrite the differential equations using dimensionless quantities given by $x = g v \, r$, $\gamma = \mu / gv$ and $\kappa = e/g$. 
After the dust has settled, the differential equations are:
\begin{align}
	\nabla^2_x \ln \! F_+^2 &= (1 + \kappa^2) F_+^2 + (1 - \kappa^2) F_-^2 - \gamma \kappa \hat{M} - \gamma \kappa \hat{N} - 2, \nonumber \\
	\nabla^2_x \ln \! F_-^2 &= (1 + \kappa^2) F_-^2 + (1 - \kappa^2) F_+^2 + \gamma \kappa \hat{M} - \gamma \kappa \hat{N} - 2, \nonumber \\
	\nabla^2_x \hat{N} &= -\gamma \left(F_+^2 + F_-^2 -2\right) + 2 \kappa^2 \hat{N} \left(F_+^2 + F_-^2\right) \nonumber \\
	&+ 2 \kappa \hat{M} \left(F_+^2 - F_-^2\right) + \gamma^2 \hat{N}, \nonumber \\
	\nabla^2_x  \hat{M} &= -\gamma \kappa \left(F_+^2 - F_-^2\right) + 2 \hat{M} \left(F_+^2 + F_-^2\right) \nonumber \\
	&+ 2 \kappa \hat{N} \left(F_+^2 - F_-^2\right) + \gamma^2 \hat{M}. 
\end{align}

The general strategy adopted here is the following: we expand the profile functions $F_+, F_-, \hat{N}, \hat{M}$ in powers of $x$ around the origin, using the generic notation $A(x) = \sum_{k} A_k x^k$. Applying these expansions in the differential equations and using the initial conditions, we can find constraints in the expansion coefficients. With these expressions at hand, we can search for numerical solutions that also satisfy the asymptotic boundary conditions using a shooting method. In general lines, for the differential equations and initial conditions considered here, there are 4 coefficients to be adjusted; the others vanish or can be determined in terms of these 4.

\begin{figure}[t!]
	\begin{minipage}[b]{0.58\linewidth}
		\includegraphics[width=\textwidth]{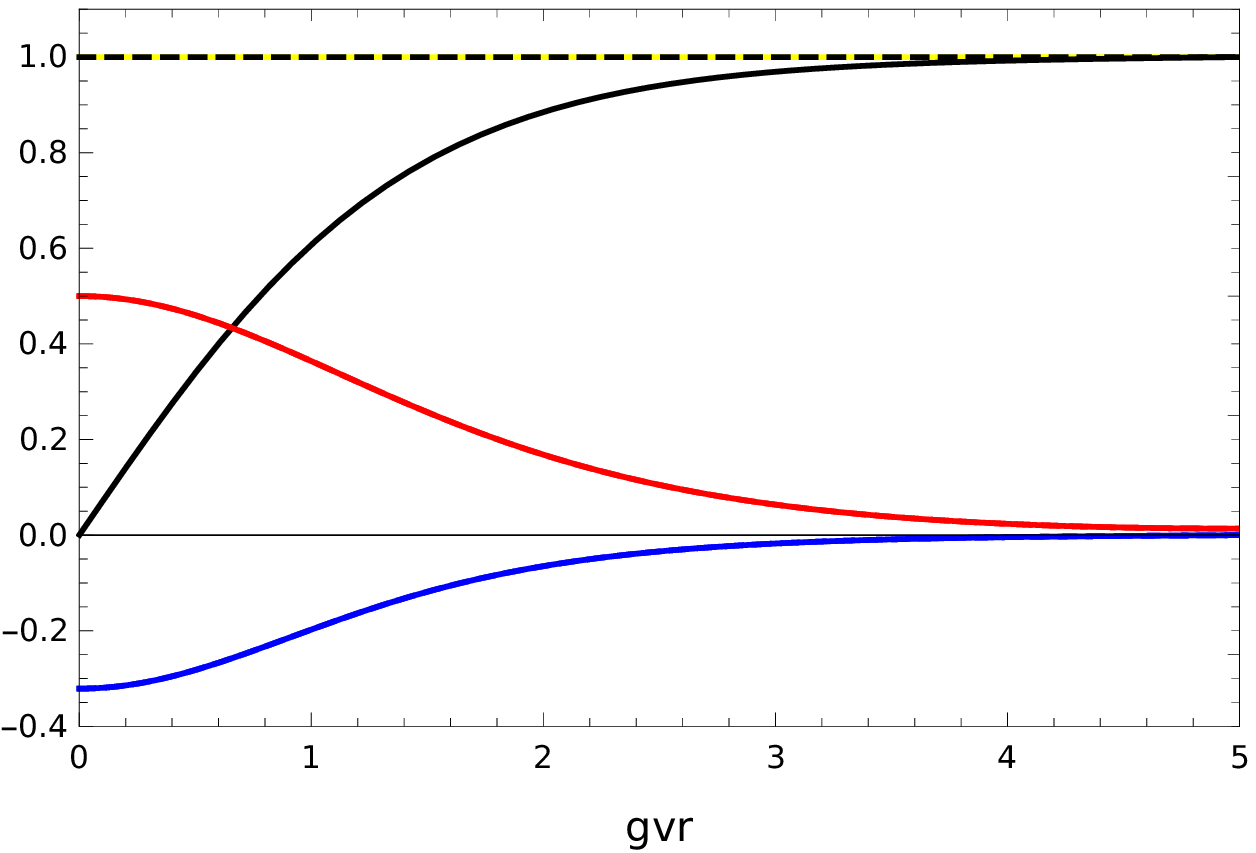}
	\end{minipage} \hfill
	\begin{minipage}[b]{0.58\linewidth}
		\includegraphics[width=\textwidth]{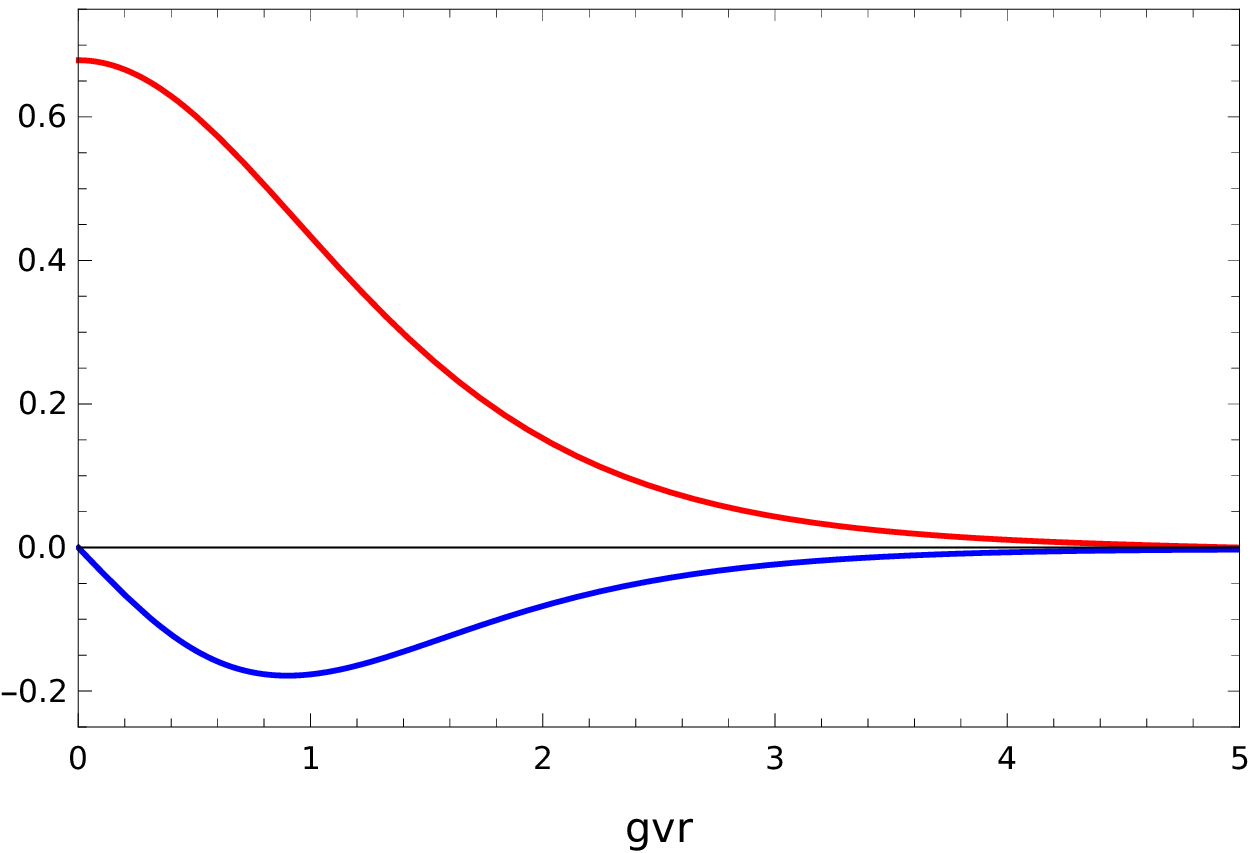}
	\end{minipage} \hfill
	\caption{ Topological vortex solution for $n=m=1/2$ and its physical fields in units of $g v^2$ as functions of $x = g v \, r$. \textit{Upper figure}: $F_+$ and $F_-$ are shown in solid and dashed black, $N=M$ in blue, and $A=a$ in red, respectively. \textit{Lower figure}: In red, the magnetic field; in blue, the electric field. Notice that in this case we have $B=b$ and $E_r = e_r$.}
	\label{solnmeiommeioT}
\end{figure}	 
\begin{figure}[t!]
	\begin{minipage}[b]{0.58\linewidth}
		\includegraphics[width=\textwidth]{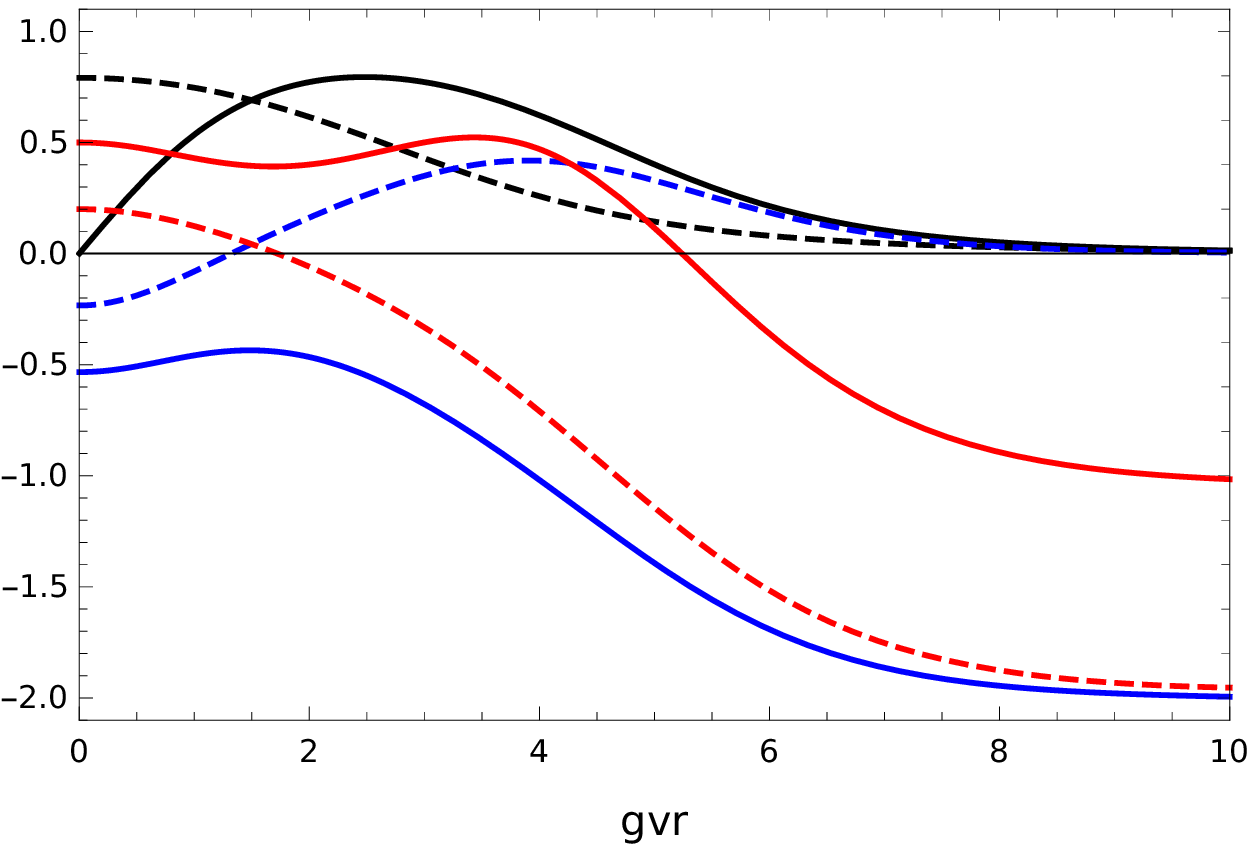}
	\end{minipage} \hfill
	\begin{minipage}[b]{0.58\linewidth}
		\includegraphics[width=\textwidth]{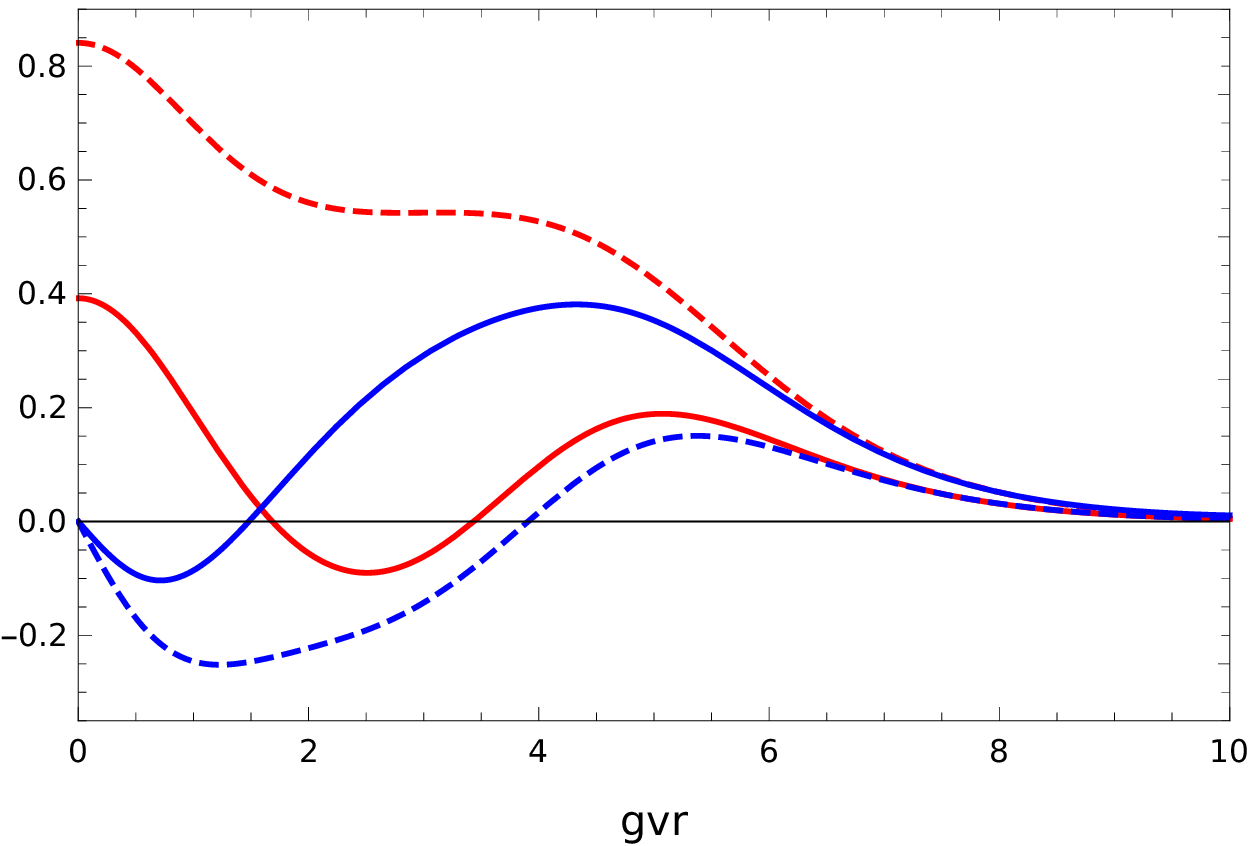}
	\end{minipage} \hfill
	\caption{Non-topological soliton for $n=1/2, m=1/2$ and its physical fields in units of $g v^2$, as functions of $x = g v \, r$. \textit{Upper figure}: $F_+$ and $F_-$ are shown in solid and dashed black, $N$ and $M$ in solid and dashed blue, $A$ and $0.4 a$ in solid and dashed red, respectively ($a$ was rescaled to facilitate the visualization). Here we have $\beta  \simeq 1.03$ and $\alpha \simeq 4.88 $. \textit{Lower figure}: The magnetic (solid red), g-magnetic (dashed red), electric (solid blue) and g-electric (dashed blue) fields.}
	\label{solnmeiommeioNT}
\end{figure}

\begin{figure}[t!]
	\begin{minipage}[b]{0.58\linewidth}
		\includegraphics[width=\textwidth]{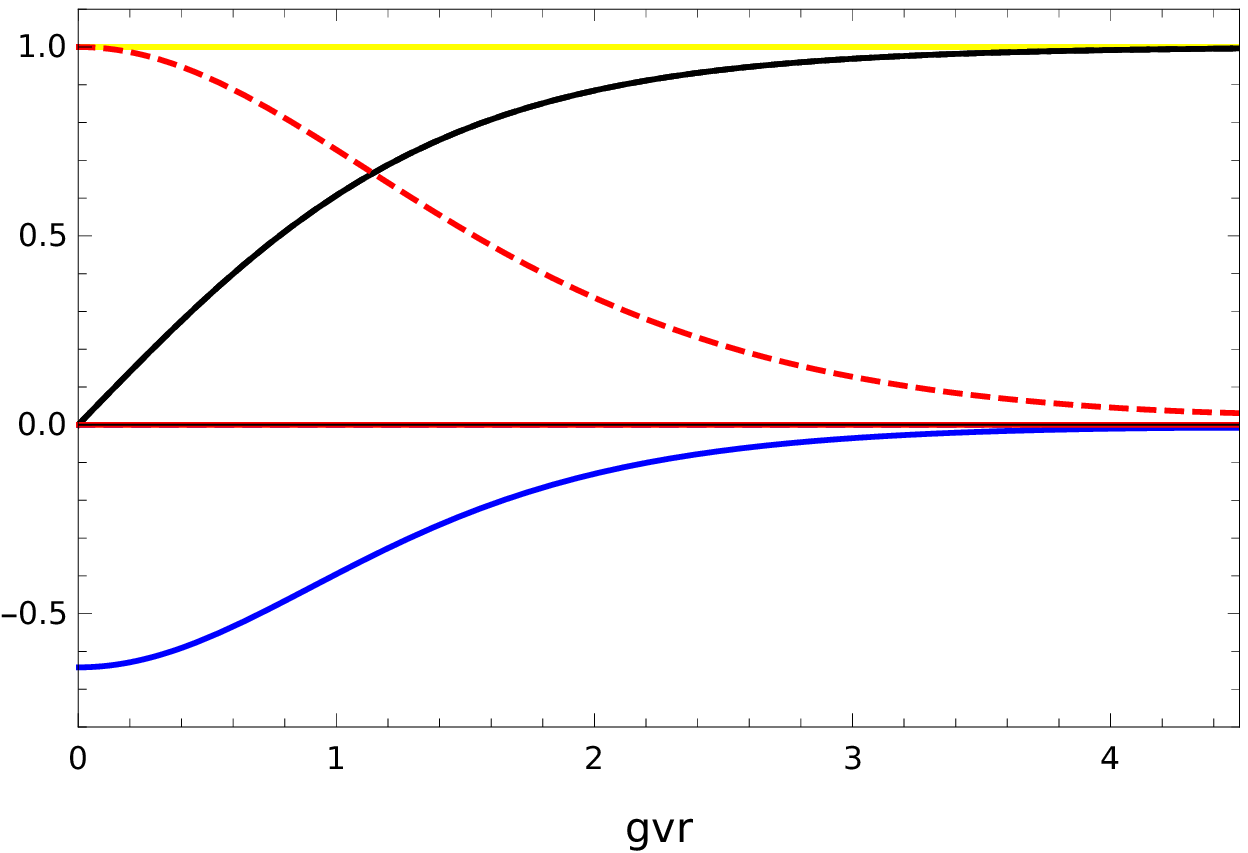}
	\end{minipage} \hfill
	\begin{minipage}[b]{0.58\linewidth}
		\includegraphics[width=\textwidth]{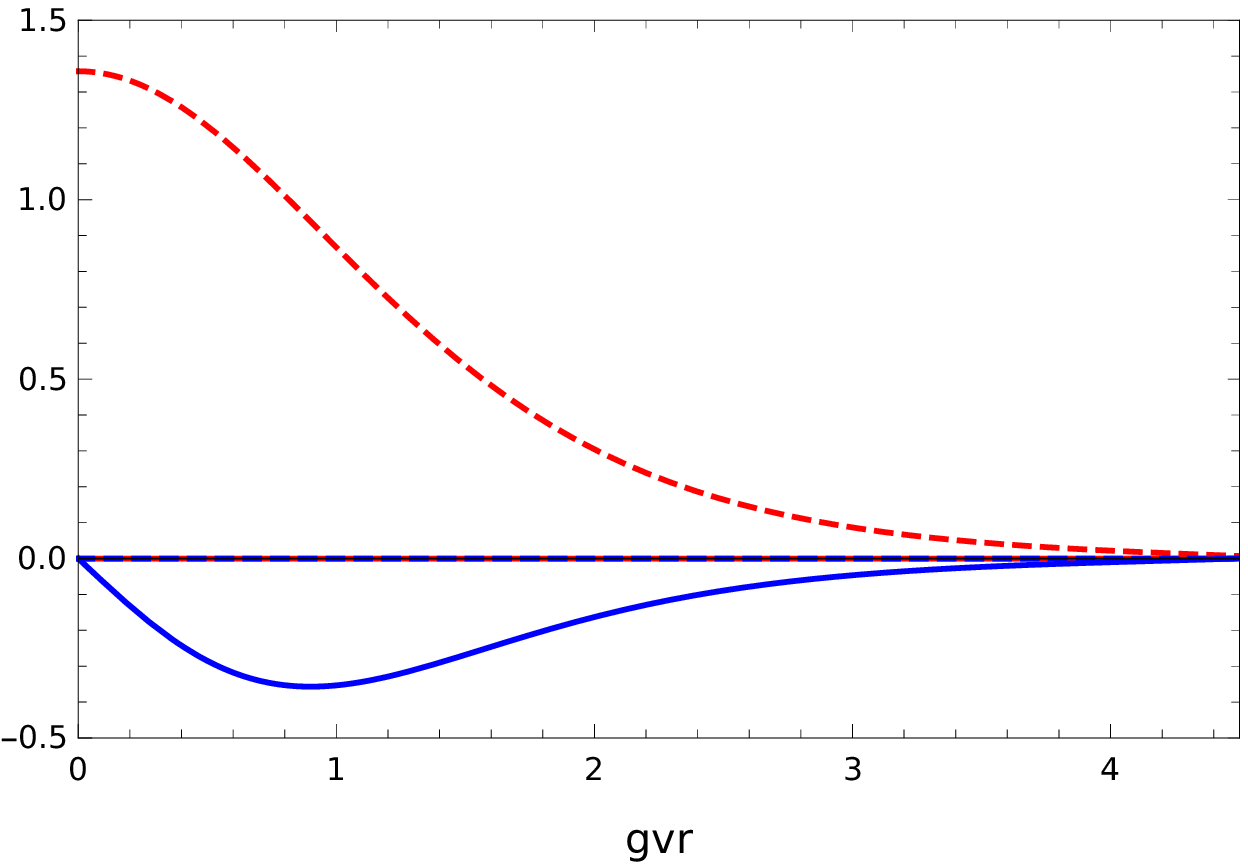}
	\end{minipage} \hfill
	\caption{Topological vortex for $n=1, m=0$ and its physical fields  in units of $g v^2$, as functions of $x = g v \, r$. \textit{Upper figure}:  $F_+ = F_-$ are shown in black, $N$ and $M$ in solid and dashed blue, $A$ and $a$ in solid and dashed red, respectively. Here we have $A = M = 0$. \textit{Lower figure}: In red, the g-magnetic field; in blue, the electric field. Here have $B = e_r = 0$.}
	\label{soln1m0T}
\end{figure}

\begin{figure}[t!]
	\begin{minipage}[b]{0.58\linewidth}
		\includegraphics[width=\textwidth]{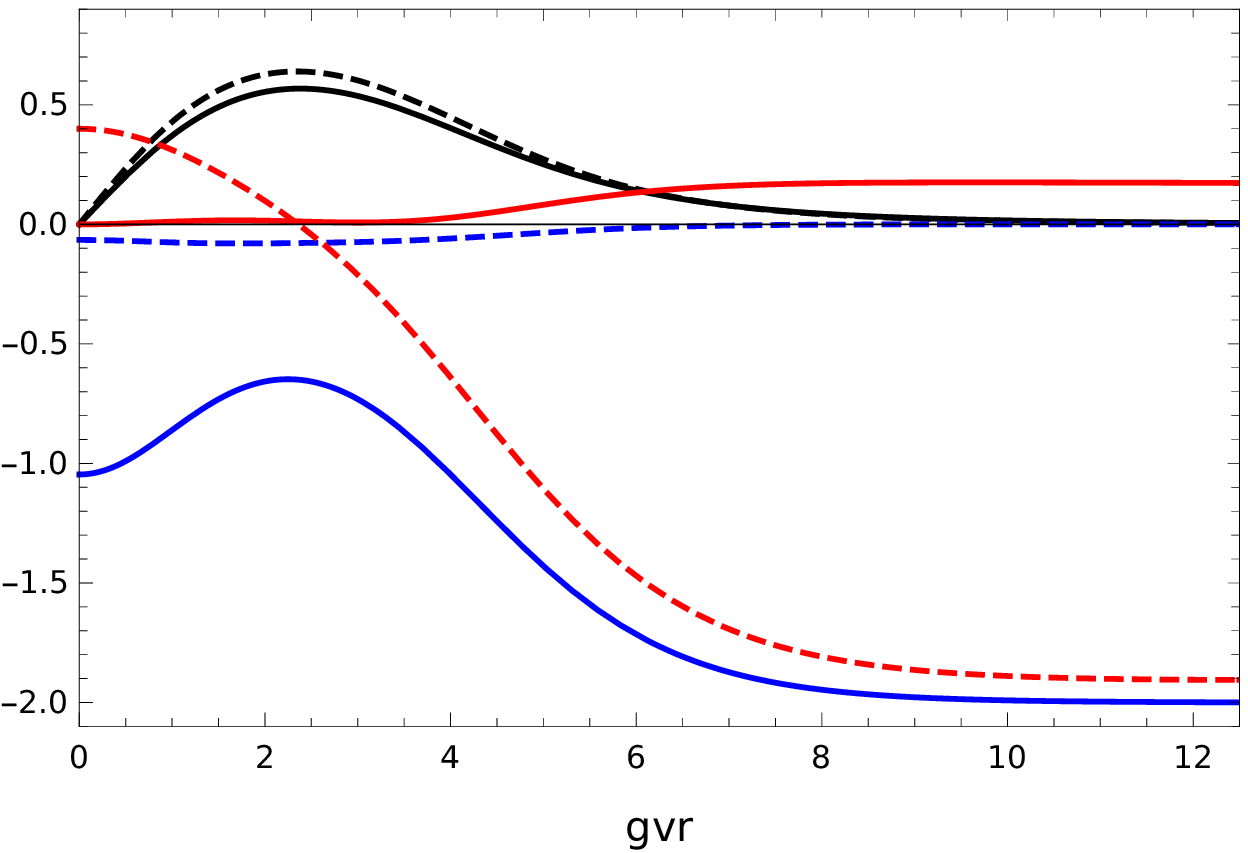}
	\end{minipage} \hfill
	\begin{minipage}[b]{0.58\linewidth}
		\includegraphics[width=\textwidth]{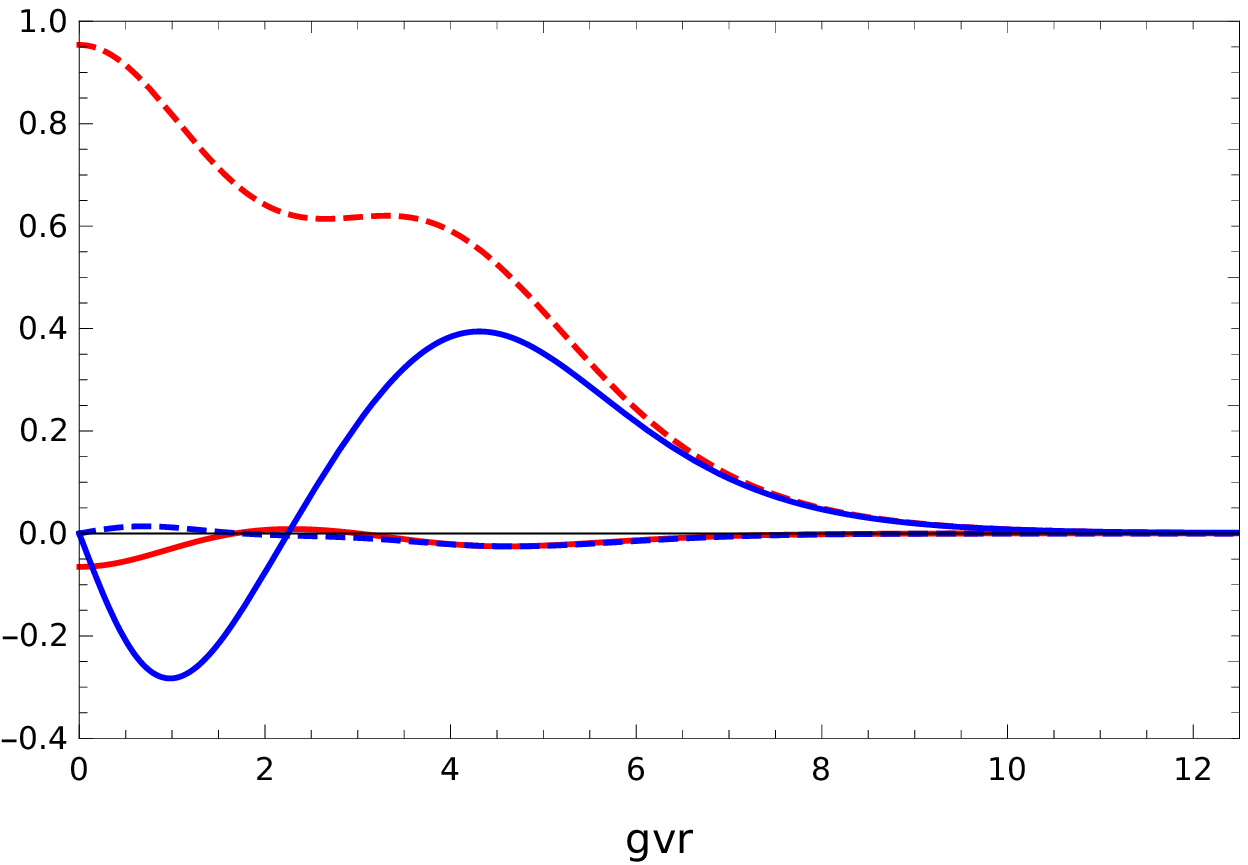}
	\end{minipage} \hfill
	\caption{Non-topological soliton for $n=1, m=0$ and its physical fields in units of $gv^2$, as functions of $x = g v \, r$. \textit{Upper figure}: $F_+$ and $F_-$ are shown in solid and dashed black, $N$ and $M$ in solid and dashed blue, $A$ and $0.4 a$ in solid and dashed red, respectively ($a$ was rescaled to facilitate the visualization). Here we have $\beta  \simeq -0.17$ and $\alpha \simeq 4.76 $. \textit{Lower figure}: The magnetic (solid red), g-magnetic (dashed red), electric (solid blue), and g-electric (dashed blue) fields.}
	\label{soln1m0NT}
\end{figure}

\begin{figure}[t!]
	\begin{minipage}[b]{0.58\linewidth}
		\includegraphics[width=\textwidth]{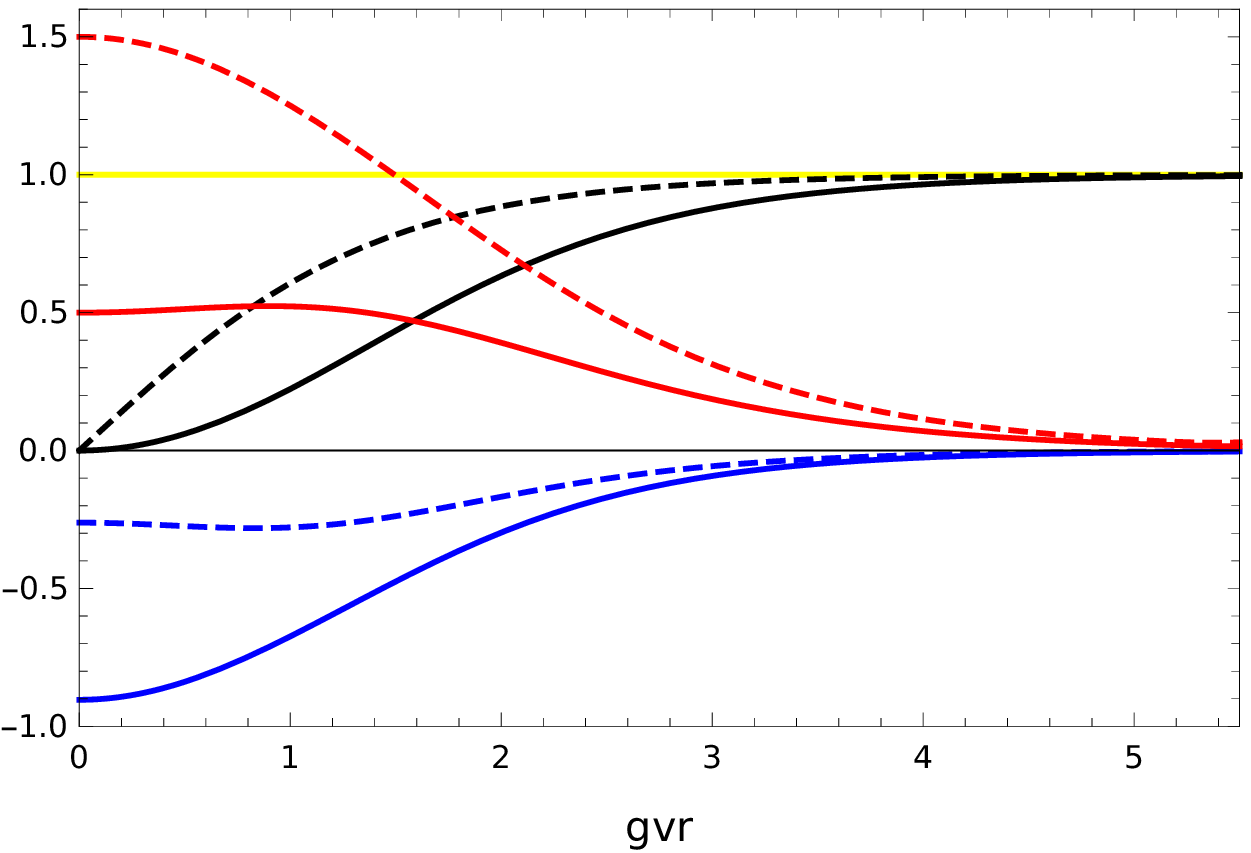}
	\end{minipage} \hfill
	\begin{minipage}[b]{0.58\linewidth}
		\includegraphics[width=\textwidth]{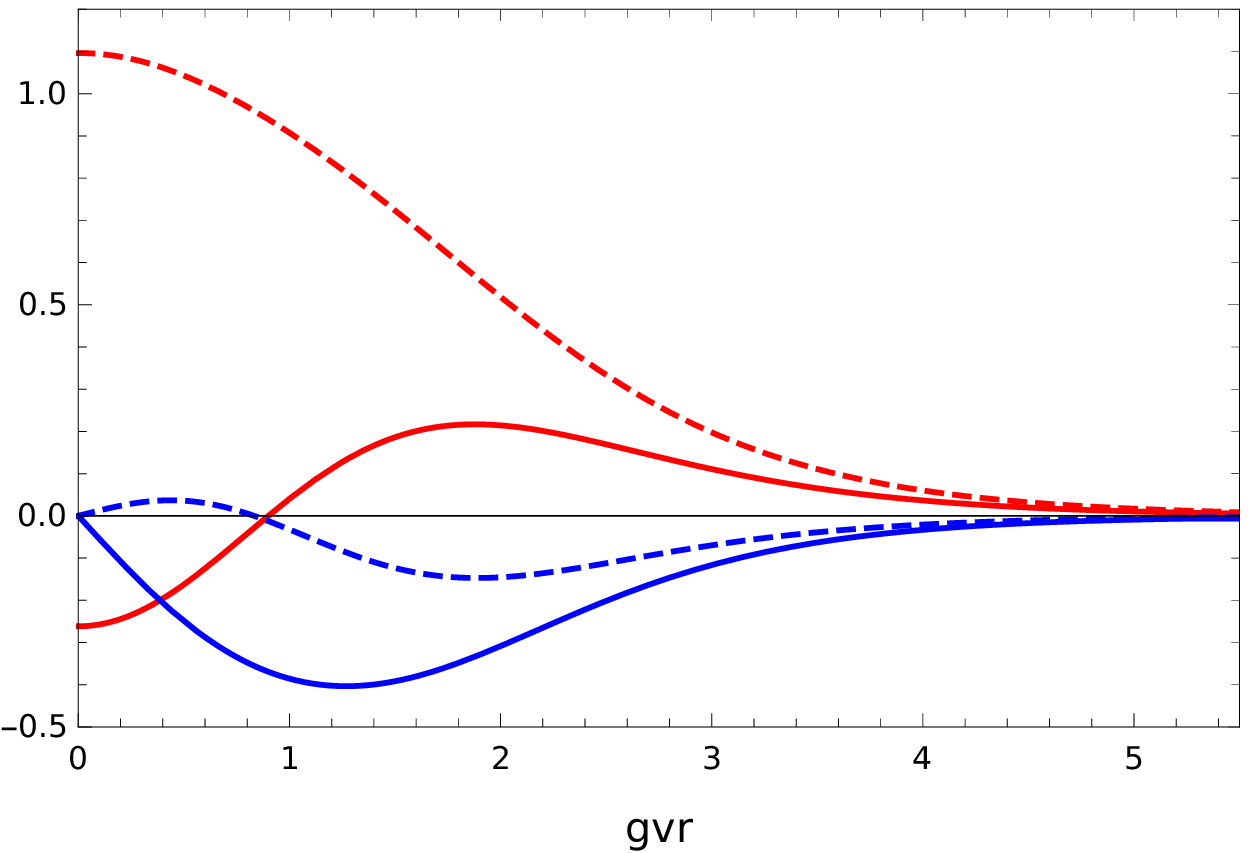}
	\end{minipage} \hfill
	\caption{Topological vortex for $n=3/2, m=1/2$ and its physical fields in units of $gv^2$, as functions of $x = g v \, r$. \textit{Upper figure}:  $F_+$ and  $F_-$ are shown in solid and dashed black, $N$ and $M$ in solid and dashed blue, $A$ and $a$ in solid and dashed red, respectively.  \textit{Lower figure}: The magnetic (solid red), g-magnetic (dashed red),  electric (solid blue), and g-electric (dashed blue) fields.}
	\label{soln3meiommeioT}
\end{figure}

\begin{figure}[t!]
	\begin{minipage}[b]{0.58\linewidth}
		\includegraphics[width=\textwidth]{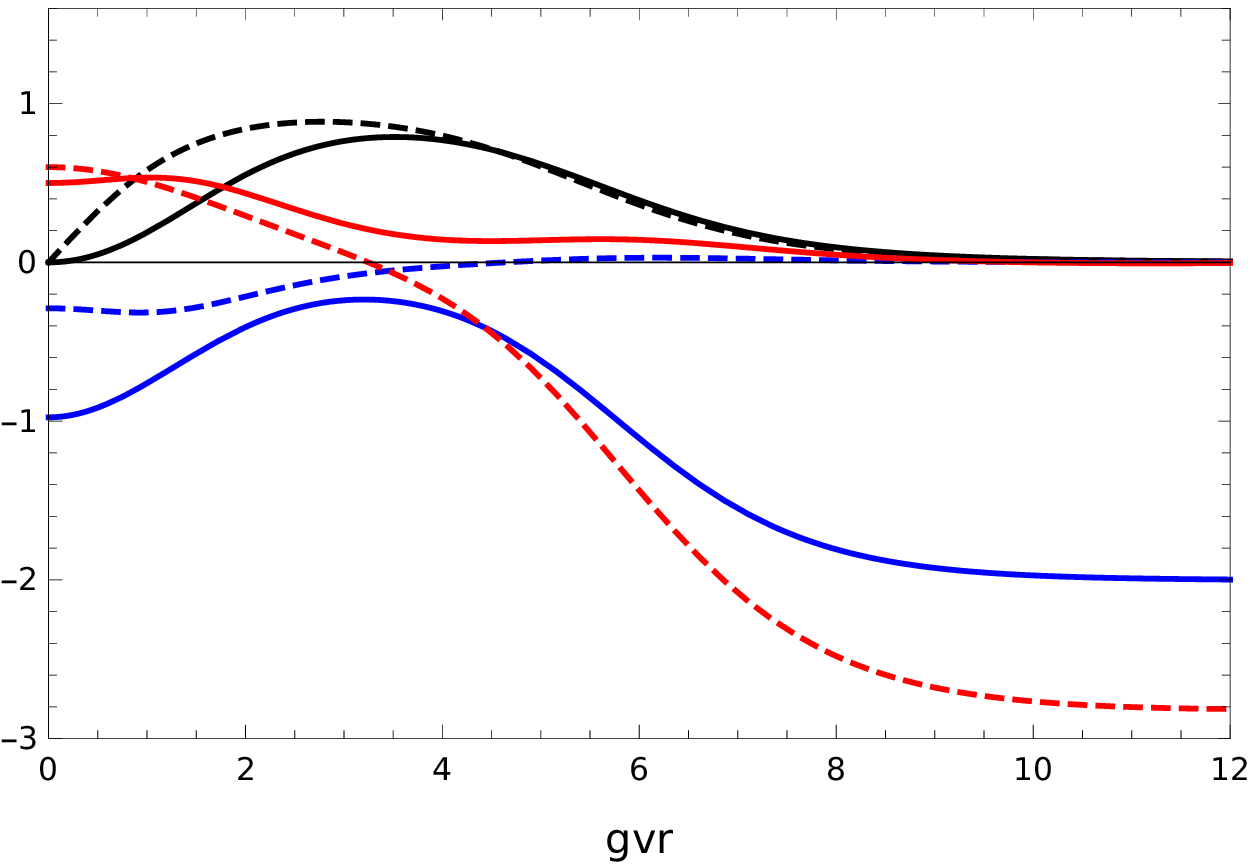}
	\end{minipage} \hfill
	\begin{minipage}[b]{0.58\linewidth}
		\includegraphics[width=\textwidth]{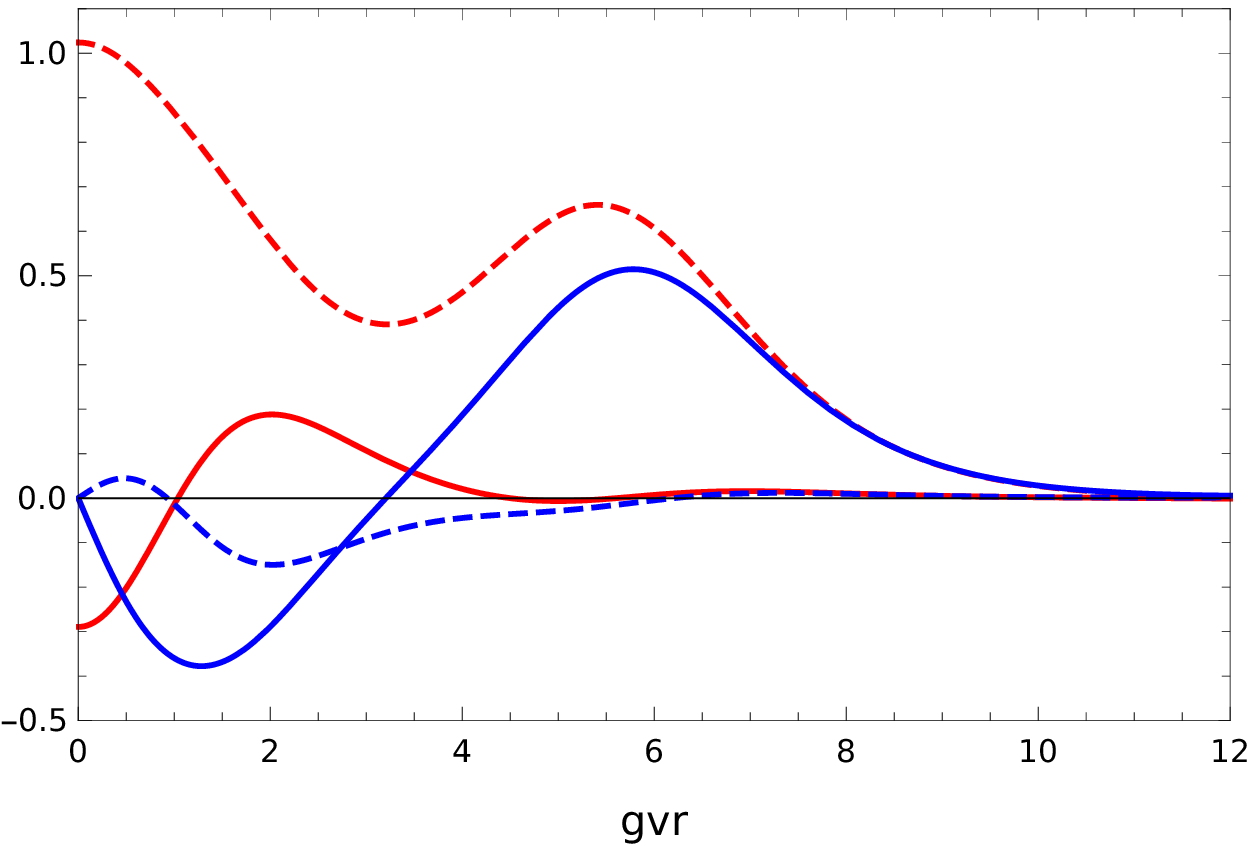}
	\end{minipage} \hfill
	\caption{Non-topological soliton for $n=3/2, m=1/2$ and its physical fields in units of $gv^2$, as functions of $x = g v \, r$. \textit{Upper figure}:  $F_+$ and  $F_-$ are shown in solid and dashed black; $N$ and $M$ in solid and dashed blue; $A$ and $0.4 a$ in solid and dashed red, respectively ($a$ was rescaled to facilitate the visualization). Here we have $\beta \simeq 0.00 $ and $\alpha \simeq 7.03$.  \textit{Lower figure}: The magnetic (solid red), g-magnetic (dashed red), electric (solid blue) and g-electric (dashed blue) fields }
	\label{soln3meiommeioNT}
\end{figure}

\begin{figure}[t!]
	\begin{minipage}[b]{0.58\linewidth}
		\includegraphics[width=\textwidth]{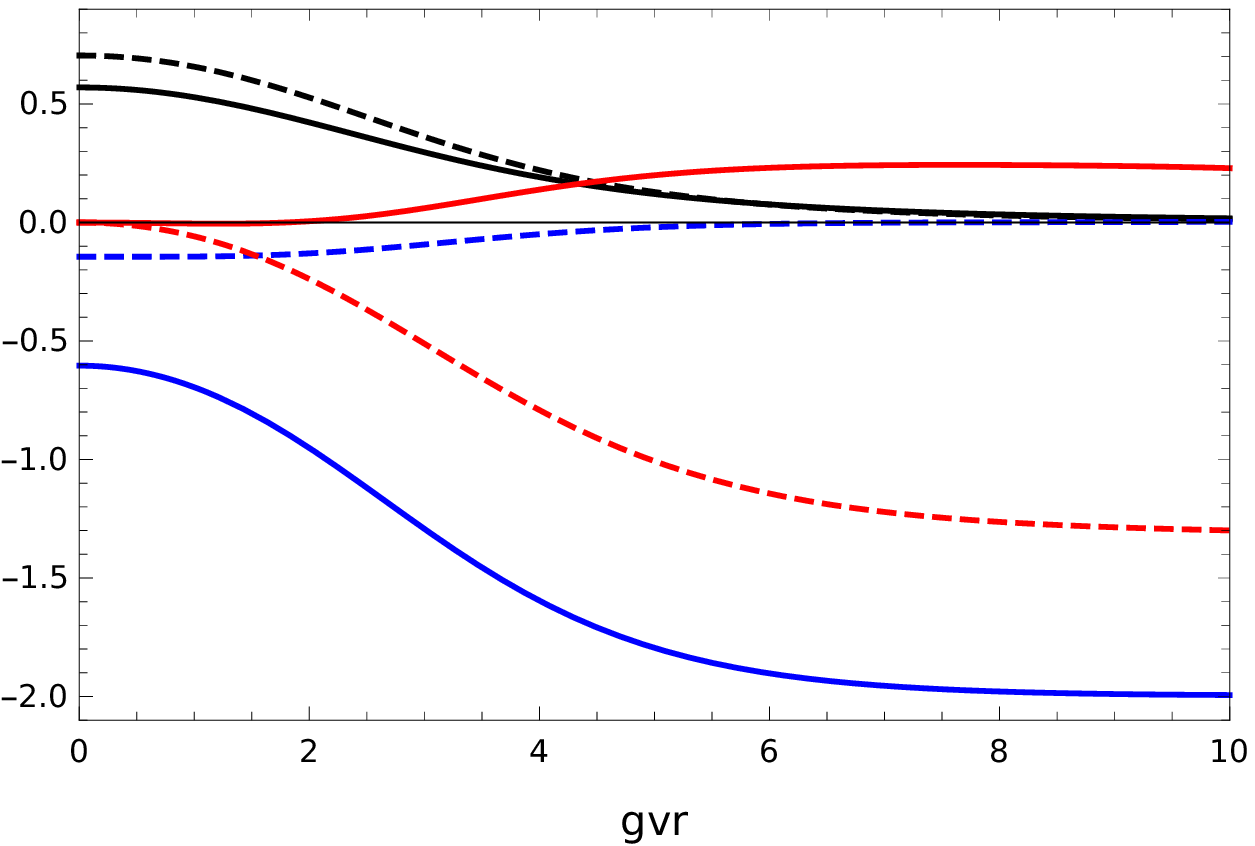}
	\end{minipage} \hfill
	\begin{minipage}[b]{0.58\linewidth}
		\includegraphics[width=\textwidth]{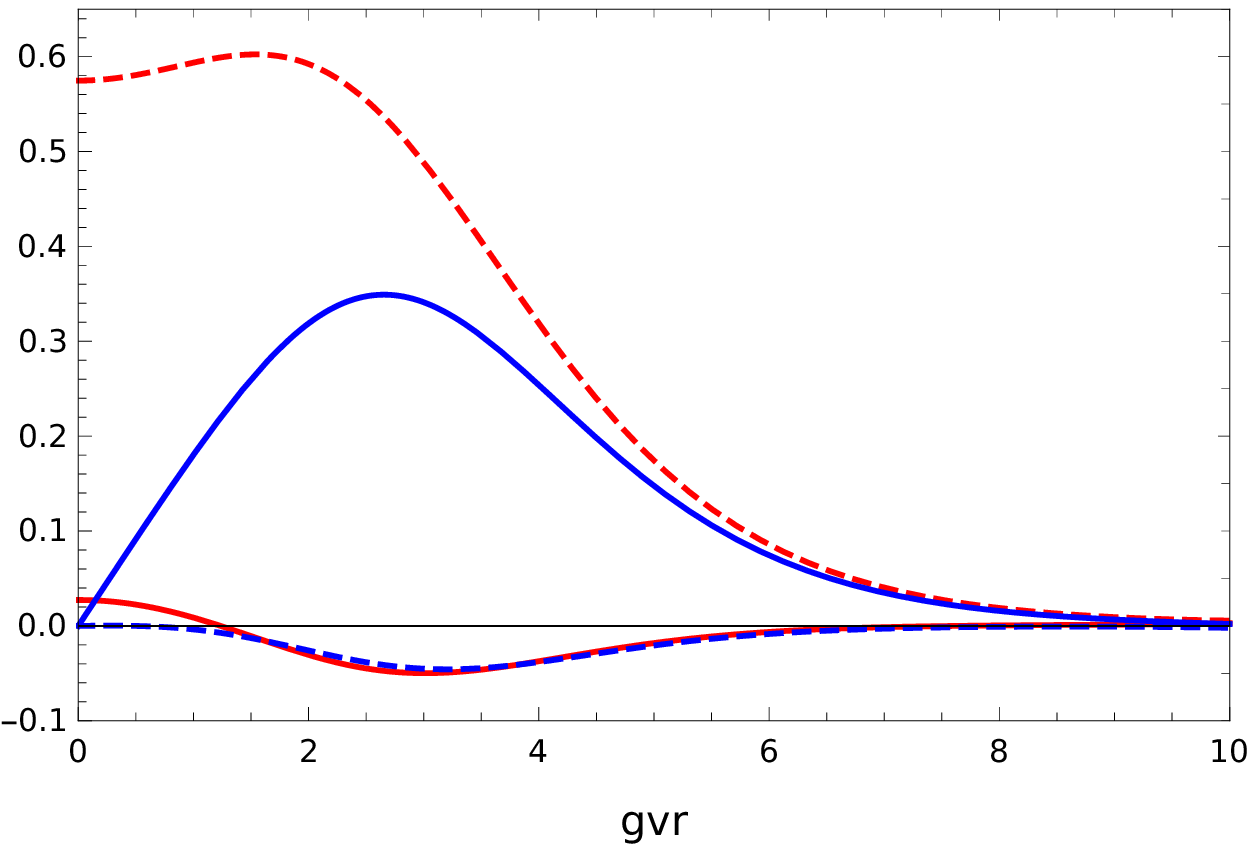}
	\end{minipage} \hfill
	\caption{Non-topological soliton for $n=m=0$ and its physical fields in units of $gv^2$, as functions of $x = g v \, r$. \textit{Upper figure}:  $F_+$ and $F_-$ are shown in solid and dashed black; $N$ and $M$ in solid and dashed blue; $A$ and $0.4 a$ in solid and dashed red, respectively ($a$ was rescaled to facilitate the visualization). Here we have $\beta  \simeq -0.23$ e $\alpha \simeq 3.25 $. \textit{Lower figure}: The magnetic (solid red), g-magnetic (dashed red), electric (solid blue) and g-electric (dashed blue) fields.}
	\label{soln0m0NT}
\end{figure}

In the following, we consider some examples with the lowest possible values for $n$ and $m$ that represent each possible class of solutions. 
The topological vortices (asymptoting to the $(1,1)$-vacuum) and non-topological solitons (asymptoting to the $(0,0)$-vacuum), with its physical fields ({\it i.e.}, electric, magnetic, g-electric and g-magnetic), for the cases $(n,m) = (\frac{1}{2}, \frac{1}{2}) , (1,0) , (\frac{3}{2}, \frac{1}{2})$ and $(0,0)$ are shown in Figs.~\ref{solnmeiommeioT}, \ref{solnmeiommeioNT}, \ref{soln1m0T}, \ref{soln1m0NT}, \ref{soln3meiommeioT}, \ref{soln3meiommeioNT}, \ref{soln0m0NT}, respectively. Their relevant physical properties are shown in Table~\ref{tableTNT}. The charges are not shown there, but can immediately be found remembering that $Q = \mu \chi$ and $G = \mu \Phi$. 
Here we adopt $\gamma = \kappa = 1$ for simplicity, but in the end of this section we comment about the relevant changes in the solutions when we vary these coefficients.

The topological vortices have quantized physical properties while non-topological solitons do not, and  the later have energy bigger than the former. 
The angular momentum for topological vortices is quantized, proportional to the product of the charges and fractional, exhibiting an anyonic nature.  
For $n=m=0$, the only solution asymptoting to the $(1,1)$-vacuum is the trivial one. 

The multiplicity of zeros of the scalar field is related to the winding number of the vortex. Therefore, the power-law behavior of $F_+$ and $F_-$ in Eq.~\eqref{behaviororiginscalar} clearly indicates that the true winding numbers are given by $ n + m $ and $ n - m$, instead of $m $ and $n$ separately, as is clearly illustrated in the explicit solutions that we found.


The most distinctive signature of a symmetry in a system is the presence of a degeneracy in the spectrum, hence it is reasonable to expect that the parity invariance of our model should reproduce this effect. To this end, we state how the vortex solutions change under parity transformations:  $\left(n, m\right) \rightarrow \left(n, -m\right)$, $r \rightarrow r, \theta \rightarrow -\theta - \pi, F_\pm \rightarrow F_\mp, M \rightarrow -M, N \rightarrow N, \beta \rightarrow -\beta, \alpha \rightarrow \alpha$, being all the others directly inferred from the self-duality equations. 

Considering the self-dual topological vortices, that is, that satisfy $ E \propto |n|$, it is immediate to conclude that a given solution and its parity-transformed version have the same energy. But the complete independence of the energy from $m$ suggests a much greater degeneracy. In fact, from the condition of regularity of the solutions as $r \rightarrow 0$, we observed that $\pm n \geq |m|$, which in turn implies that,  for $n>0$ ($n<0$) there are $2n+1$ ($2|n|+1$) solutions of the same energy. Since the energy does not depend on the sign of $n$, we obtain a $2(2|n|+1)$-fold degeneracy. It is reasonable to speculate whether this comes from a larger symmetry group. In the light of previous comments, a  good candidate would be supersymmetry  or, given the structure of the degeneracy, an internal $SU(2)$. This investigation should be pursued elsewhere. The above discussion does not apply to the non-topological solitons.

In our previous work~\cite{PippoWell}, we studied the energies of different vortices,  obtaining the following result: $M_{(1/2,1/2)} + M_{(1/2,-1/2)} = 2 M_{(1/2,1/2)} > M_{(1,0)}$, where $M_{(n,m)}$ is the mass associated with the $(n,m)$- topological vortex. 
The left-hand side of the inequality represents the static energy of well-separated $F_+$ and $F_-$ vortices of winding 1, while the right-hand side is their energy when superimposed at the origin. Therefore, the inequality suggested a possible attraction between these vortices. Now, in the self-dual model studied here, on the other hand, $2 M_{(1/2,1/2)} = M_{(1,0)}$, indicating that these vortices do not interact with each other, allowing, for example, the existence of static multi-vortex configurations, as it is usually the case for self-dual models.

\begin{figure}[t!]
	\begin{minipage}[b]{0.58\linewidth}
		\includegraphics[width=\textwidth]{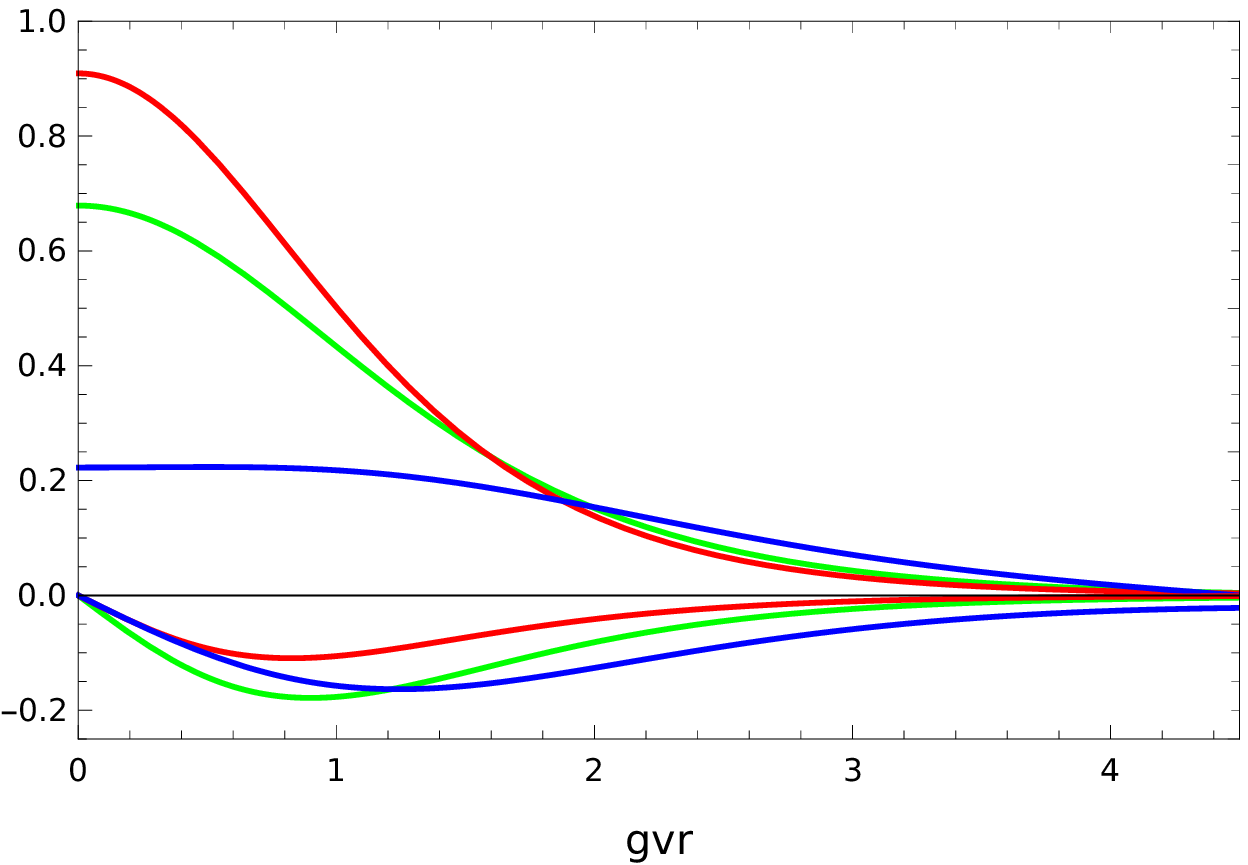}
	\end{minipage} \hfill
	\begin{minipage}[b]{0.58\linewidth}
		\includegraphics[width=\textwidth]{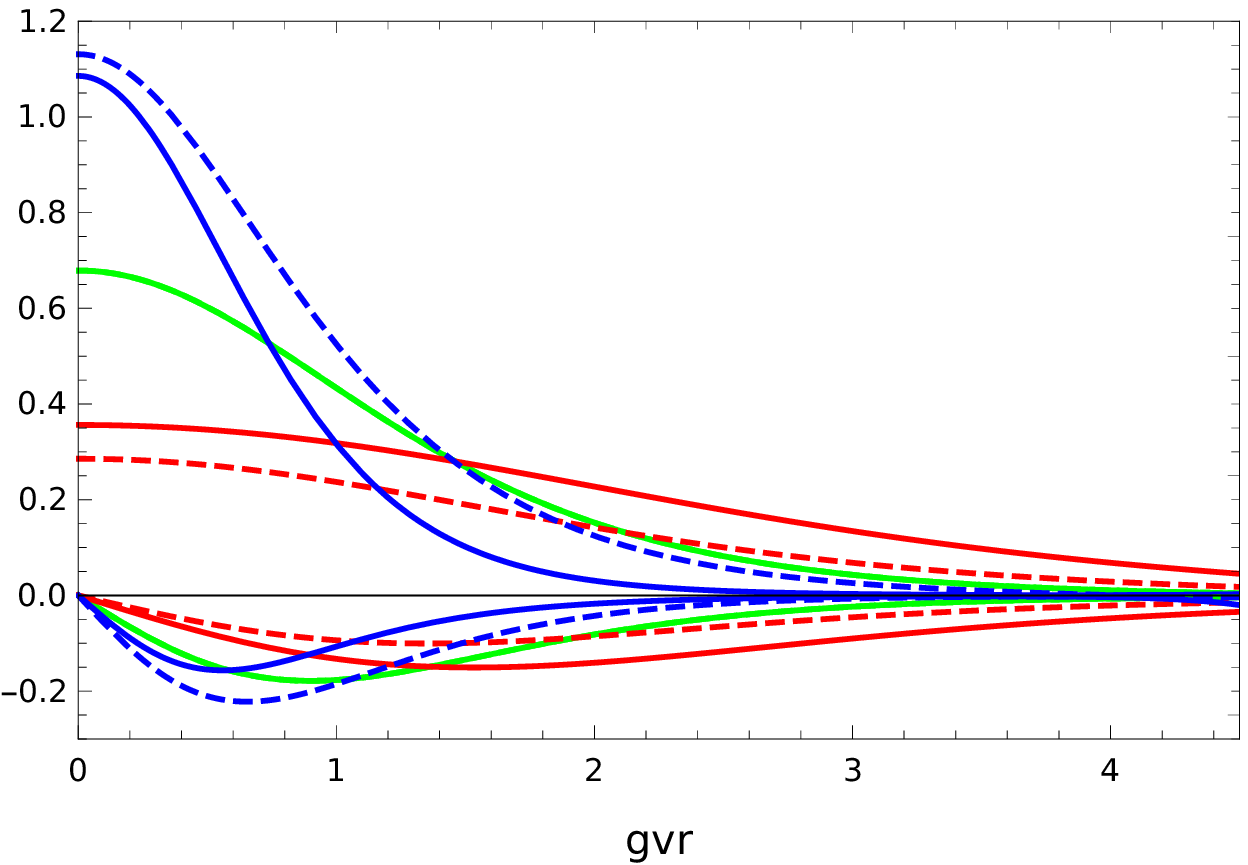}
	\end{minipage} \hfill
	\caption{ Physical fields associated with the $n=m=1/2$ topological vortex. \textit{Upper figure}: The electric (lower half-plane) and magnetic (upper half-plane) fields for $\kappa = 1$ and $\gamma = 1$ (green), $0.5$ (red), $2$ (blue). \textit{Lower figure}: The magnetic, g-magnetic, electric and g-electric fields, for $\gamma=1$ and $\kappa = 1$ (green), $0.5$ (red), $2$ (blue). The solid lines refer to $B$ and $E_r$; the dashed lines to $b$ and $e_r$. $B$ and $b$ are shown in the upper half-plane; $E_r$ and $e_r$ in the lower half-plane.}
	\label{EBnmeiommeio11vargamma}
\end{figure}

In this section we considered $\gamma = \kappa =1$ for simplicity, but the existence of solitons here is not conditioned to this assumption, and we were able to find solutions for different values of these coefficients. Interestingly enough, keeping $\kappa$ fixed and increasing $\gamma$, we see that the magnetic field at the origin decreases; decreasing $\gamma$, the magnetic field increases, cf. Fig.~\ref{EBnmeiommeio11vargamma}. Since  $\gamma \propto \mu$, this suggests that it would reach a maximum value in the pure Maxwell limit and go to zero in the pure CS limit, as it happens in the usual Maxwell-CS case~\cite{Boyanovsky}. It is well-known that in the absence of a CS term, the vortices are electrically neutral, therefore having zero electric field. In fact, we observed that in decreasing $\gamma$, the maximum value of the electric field diminished, in accordance with what is expected.
Furthermore, keeping $\gamma$ fixed and considering $\kappa \neq 1$, we can see that for  $n=m=1/2$, the electric and magnetic fields will not be degenerate anymore, cf. Fig.~\ref{EBnmeiommeio11vargamma}. In the other  examples considered, taking $\kappa \neq 1$ does not lead to significant qualitative changes.

Finally, we also found solitons asymptoting to the parity-breaking $(1,0)$- and $(0,1)$-vacua. Given the rich vacuum structure of this theory, in principle, one could also find domain walls connecting any pair of degenerate vacua. These were not discussed here for reasons of scope, since we focused on the parity-invariant cases. 
\\

%
%

\section{Concluding Remarks}\label{conclusion}

In this work, we investigated a self-dual version of the parity-invariant Maxwell-CS $U(1) \times U(1)$ model coupled with scalar matter. We obtained a Bogomol'nyi bound for the energy, whose saturation led us to first-order self-duality equations. We exhibited explicit numerical solutions corresponding to topological vortices and non-topological solitons, and discussed their main properties.	



It is important to consider the quantum aspects of this model, to study the dynamics of these vortices, and to investigate the existence of dualities in this context.
Interestingly enough, similar models can find many applications in condensed matter~\cite{Kou1, Kou2, Kou3, Kou4, Qi, Ye, Diamantini1, Diamantini2, Diamantini3, Sakhi, Oswaldo, Wellisson}, and it would be exciting to find a physical system accurately described by our model, allowing it to be experimentally realized. 

It is also important to investigate the physics of the parity and time-reversal breaking $(1,0)$- and $(0,1)$-vacua, and in particular, to study solitons asymptoting to them, as well as the existence of domain walls connecting the degenerate phases of this model. 
The role of parity and time-reversal in superconductors is a topic that has been attracting much interest recently~\cite{ExpRecente1, ExpRecente2, ExpRecente3}, and perhaps our model could find some use in this subject.

A natural development is the supersymmetric extension of this model, since self-duality and supersymmetry are intimately related~\cite{Susy1,Susy2,Susy3,Susy4}. Moreover, supersymmetry can find applications in graphene~\cite{Helayel, EzawaSusy, Cabrera} and can dynamically emerge in condensed matter systems~\cite{Lee2, Grover, Jian, Ponte}. Therefore, this could lead to interesting physical results. 


\begin{acknowledgments}
The authors are grateful to J. A. Helay\"el-Neto for helpful discussions, and to P. C. Malta for useful comments.
The authors also thank the Brazilian scientific support agencies CNPq and FAPERJ for financial support.	
\end{acknowledgments}
	
\newpage



\begin{thebibliography}{99}


\bibitem{NielsenOlesen}
H. B. Nielsen and P. Olesen, {\it Vortex-line models for dual strings}, Nucl. Phys. B {\bf 61} 45 (1973).
				
\bibitem{Abrikosov}
A. A. Abrikosov, {\it On the Magnetic properties of superconductors of the second group}, Sov. Phys. JETP {\bf 5}, 1174 (1957).	

\bibitem{GinzburgLandau}
V. L. Ginzburg  and L. D. Landau,   Zh. Eksp. Teor. Fiz. {\bf 20}, 1064 (1950).

\bibitem{Bogomol'nyi}
E. B. Bogomolny, {\it Stability of Classical Solutions }, Sov. J. Nucl. Phys. {\bf 24}, 449 (1976).

\bibitem{VegaSchaposnik}
H. J. de Vega and F. A. Schaposnik, {\it Classical vortex solution of the Abelian Higgs model}, Phys. Rev. D {\bf 14}, 110 (1976).		

\bibitem{DeserJackiwTempleton1}
S. Deser, R. Jackiw, and S. Templeton, {\it Three-Dimensional Massive Gauge Theories}, Phys. Rev. Lett.	{\bf 48},  975 (1982).

\bibitem{DeserJackiwTempleton2}
S. Deser, R. Jackiw, and S. Templeton, {\it Topologically Massive Gauge Theories},  Ann. Phys. {\bf 140},  372 (1982); Ann. Phys. {\bf 281},  409 (2000).

\bibitem{Schonfeld}
J. F. Schonfeld, {\it A mass term for three-dimensional gauge fields}, Nucl. Phys. B {\bf 185}, 157 (1981).


\bibitem{ChernSimons}
S.-S. Chern and J. Simons, {\it  Characteristic Forms and Geometric Invariants}, Ann. Math. {\bf 99}, 48 (1974).

\bibitem{Dunne}
G. V. Dunne, {\it Aspects of Chern-Simons theory}, arXiv:9902115.

\bibitem{PaulKhare}
S. K. Paul and A. Khare, {\it Charged Vortices in Abelian Higgs Model with Chern-Simons Term}, Phys. Lett. B {\bf 174}, 420 (1986).

\bibitem{JatkarKhare}
D. P. Jatkar and A. Khare, {\it Peculiar charged vortices in Higgs models with pure Chern-Simons term}, Phys. Lett B {\bf 236}, 283 (1990).


\bibitem{HongKimPac}
J. Hong, Y. Kim, and P. Y. Pac, {\it Multivortex solutions of the Abelian Chern-Simons-Higgs theory}, Phys. Rev. Lett. {\bf 64}, 2230 (1990).

\bibitem{JackiwWeinberg}	
R. Jackiw and E. Weinberg, {\it Self-dual Chern-Simons vortices}, Phys. Rev. Lett. {\bf 64}, 2234 (1990).

\bibitem{JackiwLeeWeinberg}
R. Jackiw, K. Lee and E. Weinberg, {\it Self-dual Chern-Simons solitons}, Phys. Rev. D {\bf 42}, 3488 (1990).

\bibitem{LeeLeeMin}
C. Lee, K. Lee, and H. Min,  {\it Self-dual Maxwell Chern-Simons solitons}, Phys. Lett. B {\bf 252}, 79 (1990).

\bibitem{Kim2}
C. Kim, {\it Self-dual vortices in the generalized Abelian Higgs model with independent Chern-Simons interaction}, Phys. Rev. D {\bf 47}, 673 (1993).

\bibitem{Rim}
C. Lee, H. Min, and C. Rim, {\it Zero modes of the self-dual Maxwell Chem-Simons solitons}, Phys. Rev. D {\bf 43}, 4100 (1991).

\bibitem{Susy1}
E. Witten and D. Olive, {\it Supersymmetry algebras that include topological charges}, Phys. Lett. B {\bf 78},  97 (1978).

\bibitem{Susy2}
C. Lee, K. Lee, and E. Weinberg,  {\it Supersymmetry and self-dual Chern-Simons systems}, Phys. Lett. B {\bf 243}, 105 (1990).

\bibitem{Susy3}
C. Lee, K. Lee, and H. Min, {\it Supersymmetric Chern-Simons vortex systems and fermion zero modes}, Phys. Rev. D {\bf 45}, 4588 (1992).

\bibitem{Susy4}
H. R. Christiansen, M. S. Cunha, J. A. Helay\"el-Neto, L. R. U. Manssur, and A. L. M. A. Nogueira, {\it Selfdual vortices in a Maxwell-Chern-Simons model with nonminimal coupling}, Int. J. Mod. Phys. A {\bf 14},  1721 (1999).

\bibitem{Susy5}
B.-H. Lee and H. Min, {\it Quantum aspects of supersymmetric Maxwell Chern-Simons solitons}, Phys. Rev. D {\bf 51}, 4458 (1995).

\bibitem{Susy6}
P. Arias, E. Ireson, F. A. Schaposnik, and G. Tallarita, {\it Chern–Simons–Higgs theory with visible and hidden sectors and its $\mathcal{N}=2$ SUSY extension}, Phys. Lett. B {\bf 749},  368 (2015).



\bibitem{Dunne2}
G. V. Dunne, {\it Selfdual Chern-Simons theories},  Lect. Notes Phys. M {\bf 36}, 1 (1995).

\bibitem{Horvathy1}
P. Horváthy and P. Zhang, {\it Vortices in (Abelian) Chern-Simons gauge theory}, Phys. Rept. {\bf 481},  83 (2009).

\bibitem{Hagen}
C. R. Hagen, {\it Parity conservation in Chern-Simons theories and the anyon interpretation}, Phys. Rev. Lett. {\bf 68}, 3821 (1992).

\bibitem{Wilczek}
F. Wilczek, {\it Disassembling Anyons}, Phys. Rev. Lett. {\bf 69}, 132 (1992).

\bibitem{Keifl}
R. F. Keifl {\it et al.}, {\it Search for anomalous internal magnetic fields in high-$T_c$ superconductors as evidence for broken time-reversal symmetry}, Phys. Rev. Lett. {\bf 64}, 2082 (1990).

\bibitem{Spielman}
S. Spielman {\it et al.}, {\it Test for nonreciprocal circular birefringence in $YBa_2Cu_3O_7$ thin films as evidence for broken time-reversal symmetry}, Phys. Rev. Lett. {\bf 65}, 123 (1990).

\bibitem{Lyons}
K. Lyons {\it et al.}, {\it Search for circular dichroism in high-$T_c$	superconductors}, Phys. Rev. Lett. {\bf 64}, 2949 (1990).

\bibitem{Semenoff}
G. W. Semenoff and N. Weiss, {\it 3D field theory model of a parity invariant anyonic superconductor}, Phys. Lett. B {\bf 250}, 117 (1990). 

\bibitem{Mavromatos}
N. Dorey and N. E. Mavromatos, {\it Superconductivity in 2+1 dimensions without parity or time-reversal violation}, Phys. Lett. B {\bf 250},  107 (1990). 


\bibitem{Kovner}
A. Kovner and B. Rosenstein, {\it Kosterlitz-Thouless mechanism of two-dimensional superconductivity}, Phys. Rev. B {\bf 42}, 4748 (1990).

\bibitem{Mavromatos2}
N. Dorey and N. E. Mavromatos, {\it QED3 and two-dimensional superconductivity without parity violation}, Nucl. Phys. B {\bf 386}, 614 (1992).


\bibitem{Kim}
C. Kim, C. Lee, P. Ko, B.-H. Lee, and H. Min, {\it Schrodinger fields on the plane with $[U(1)]^N$ Chern-Simons interactions and generalized selfdual solitons}, Phys. Rev. D {\bf 48}, 1821 (1993).

\bibitem{Diziarmaga1}
J. Diziarmaga, {\it Low-energy dynamics of $[U(1)]^N$ Chern-Simons solitons},  Phys. Rev. D {\bf 49}, 5469 (1994).

\bibitem{Diziarmaga2}
J. Diziarmaga, {\it Only hybrid anyons can exist in broken symmetry phase of nonrelativistic $[U(1)]^2$ Chern-Simons theory}, Phys. Rev. D {\bf 50}, R2376(R) (1994).

\bibitem{Shin1}
J. Shin and J. Yee, {\it Vortex solutions of parity invariant Chern-Simons gauge theory coupled to fermions},  Phys. Rev. D {\bf 50},  4223 (1994).

\bibitem{Shin2}
J. Shin, S. Hyun, and J. Yee, {\it Mutual fractional statistics of relativistic Chern-Simons solitons}, Phys. Rev. D {\bf 52},  2591 (1995).


%
%

\bibitem{PippoWell} 
W. B. De Lima and P. De Fabritiis, {\it Vortices in a parity-invariant Maxwell-Chern-Simons model}, arXiv: 2205.10427.


\bibitem{Kou1}
S.-P. Kou, X.-L. Qi, and Z.-Y. Weng, {\it Mutual Chern-Simons effective theory of doped antiferromagnets}, Phys. Rev. B {\bf 71}, 235102 (2005).

\bibitem{Kou2}
S.-P. Kou, M. Levin, and X.-G. Wen, {\it Mutual Chern-Simons theory for $Z_2$ topological order}, Phys. Rev. B {\bf 78}, 155134 (2008).

\bibitem{Kou3}
S.-P. Kou, X.-L. Qi, and Z.-Y. Weng, {\it Spin Hall effect in a doped Mott insulator}, Phys. Rev. B {\bf 72}, 165114 (2005).

\bibitem{Kou4}
S.-P. Kou, J. Yu, and X.-G. Wen, {\it Mutual Chern-Simons Landau-Ginzburg theory for continuous quantum phase transition of $Z_2$ topological order}, Phys. Rev. B {\bf 80}, 125101 (2009).

\bibitem{Qi}
X.-L. Qi and Z.-Y. Weng, {\it Mutual Chern-Simons gauge theory of spontaneous vortex phase}, Phys. Rev. B {\bf 76}, 104502 (2007).

\bibitem{Ye}
P. Ye, L. Zhang, and Z.-Y. Weng, {\it Superconductivity in mutual Chern-Simons gauge theory}, Phys. Rev. B {\bf 85}, 205142 (2012).


\bibitem{Diamantini1}
M. C. Diamantini, P. Sodano, and C. A. Trugenberger, {\it Self-duality and oblique confinement in planar gauge theories}, Nucl. Phys. B {\bf 448},  505 (1995).

\bibitem{Diamantini2}
M. C. Diamantini, P. Sodano, and C. A. Trugenberger, {\it Gauge theories of Josephson junction arrays}, Nucl. Phys. B {\bf 474},  641 (1996).

\bibitem{Diamantini3}
M. C. Diamantini, P. Sodano, and C. A. Trugenberger, {\it Superconductors with topological order}, Eur. Phys. J. B {\bf 53}, 19 (2006).

\bibitem{Sakhi}
S. Sakhi, {\it Tricritical behavior in the Chern-Simons-Ginzburg-Landau theory of self-dual Josephson junction arrays}, Phys. Rev. D {\bf 97}, 096015 (2018). 

%

\bibitem{Oswaldo}
O. M. Del Cima and E. S. Miranda, {\it Electron-polaron—electron-polaron bound states in mass-gap graphene-like planar quantum electrodynamics: s-wave bipolarons}, Eur. Phys. J. B {\bf 91}, 212 (2018).

\bibitem{Wellisson}
W. B. De Lima, O. M. Del Cima and E. S. Miranda, {\it On the electron–polaron–electron–polaron scattering and Landau levels in pristine graphene-like quantum electrodynamics}, Eur. Phys. J. B {\bf 93}, 187 (2020).

%
%


%
%
%


\bibitem{Boyanovsky}
D. Boyanovsky, {\it Vortices in Landau-Ginzburg theories of anyonic superconductivity}, Nucl. Phys. B {\bf 350}, 906 (1991).

\bibitem{ExpRecente1}
S. Kanasugi and Y. Yanase, {\it Anapole superconductivity from $\mathcal{P} \mathcal{T}$-symmetric mixed-parity interband pairing}, Commun Phys {\bf 5}, 39 (2022).

\bibitem{ExpRecente2}
S. K. Ghosh {\it et al.}, {\it Time-reversal symmetry breaking superconductivity in three-dimensional Dirac semimetallic silicides}, Phys. Rev. Research {\bf 4}, L012031 (2022).

\bibitem{ExpRecente3}
S. K. Ghosh {\it et al.}, {\it Recent progress on superconductors with time-reversal symmetry breaking}, J. Phys.: Condens. Matter {\bf 33}, 033001 (2021).

\bibitem{Helayel}
E. M. C. Abreu, M. A. De Andrade, L. P. G. De Assis, J. A. Helayël-Neto, A. L. M. A. Nogueira, and R. C. Paschoal , {\it A supersymmetric model for graphene}, J. High Energ. Phys. {\bf 2011}, 1 (2011).

\bibitem{EzawaSusy}
M. Ezawa, {\it Supersymmetric structure of quantum Hall effects in graphene}, Phys. Lett. A {\bf 372},  924 (2008).

\bibitem{Cabrera}
C. A. Dartora and G. G. Cabrera, {\it Wess–Zumino supersymmetric phase and superconductivity in graphene}, Phys. Lett. A {\bf 377},  907 (2013).


\bibitem{Lee2}
S.-S. Lee, {\it Emergence of supersymmetry at a critical point of a lattice model}, Phys. Rev. B {\bf 76}, 075103 (2007).

\bibitem{Grover}
T. Grover, D. N. Sheng, and A. Vishwanath, {\it Emergent Space-Time Supersymmetry at the Boundary of a Topological Phase}, Science {\bf 344}, 280 (2014).

\bibitem{Jian}
S.-K. Jian, Y.-F. Jiang, and H. Yao, {\it Emergent Spacetime Supersymmetry in 3D Weyl Semimetals and 2D Dirac Semimetals}, Phys. Rev. Lett. {\bf 114}, 237001 (2015).

\bibitem{Ponte}
P. Ponte and S.-S. Lee, {\it Emergence of supersymmetry on the surface of three-dimensional topological insulators}, New J. Phys. {\bf 16}, 013044 (2014).



%
%


%
%
%
%
%






\end{thebibliography}
\end{document}